\documentclass[pas,manuscript,showpacs,showkeys,superscriptaddress,nofootinbib]{revtex4-1}
\usepackage{amsmath, amssymb, amsthm, graphicx, epsfig, fancyhdr,epsfig, slashed, mathrsfs}

\usepackage{tikzsymbols}
\usepackage{natbib}
\usepackage{float}
\usepackage{graphicx}
\usepackage{amsmath}
\usepackage{amsfonts}
\usepackage{amssymb,ulem}
\usepackage[]{hyperref}
  \hypersetup{
  bookmarks=true,         
  unicode=true,         
  pdftoolbar=true,     
  pdfmenubar=true,     
  pdffitwindow=true,    
  pdfstartview={FitH},    
  pdfsubject={Sterilenus@DUNE},  
  pdfnewwindow=true,     
  pdfcreator={RevTeX},
  colorlinks=true,     
  linkcolor=red,   
  citecolor=blue,    
  filecolor=black,  
  urlcolor=blue,           
  }

\usepackage{hypcap}
\usepackage{subfigure}
\usepackage{slashed}
 \usepackage{setspace} 
\usepackage{caption}
\definecolor{lime}{HTML}{A6CE39}
\DeclareRobustCommand{\orcidicon}{
	\begin{tikzpicture}
		\draw[lime, fill=lime] (0,0) 
		circle [radius=0.2] 
		node[white] {{\fontfamily{qag}\selectfont \tiny ID}};
		\draw[white, fill=white] (-0.0625,0.095) 
		circle [radius=0.007];
	\end{tikzpicture}
	\hspace{-2mm}
}

\foreach \x in {A, ..., Z}{\expandafter\xdef\csname orcid\x\endcsname{\noexpand\href{https://orcid.org/\csname orcidauthor\x\endcsname}
		{\noexpand\orcidicon}}
}

\usepackage{comment}
\newcommand{\be}{\begin{equation}}
	\newcommand{\ee}{\end{equation}}
\newcommand{\bea}{\begin{eqnarray}}
	\newcommand{\eea}{\end{eqnarray}}

\newcommand{\beq}{\begin{equation}}
	\newcommand{\eeq}{\end{equation}}

\def\nn{\nonumber}

\usepackage[utf8]{inputenc}

\DeclareUnicodeCharacter{2212}{-}

\begin{document}



\title{ Exploring the Dark Sector of the inspired FNSM at the LHC}

\author{\small Amit Chakraborty \orcidA{}}
\email{amit.c@srmap.edu.in }
\affiliation{\small Department of Physics, School of Engineering and Sciences, SRM University AP, Amaravati, Mangalagiri 522240, India.}

\author{\small Dilip Kumar Ghosh  \orcidB{} }
\email{tpdkg@iacs.res.in} 
\affiliation{\small School of Physical Sciences, Indian Association for 
the Cultivation of Science, 2A $\&$ 2B Raja S.C. Mullick Road, 
Kolkata 700032, India.}

\author{\small Najimuddin Khan \orcidC{}}
\email{nkhan.ph@amu.ac.in}
\affiliation{\small Department of Physics, Aligarh Muslim University, Aligarh-202002, India\vspace{0.10cm}}

\author{\small Stefano Moretti  \orcidD{}}
\email{s.moretti@soton.ac.uk; stefano.moretti@physics.uu.se }
\affiliation{\small   School of Physics $\&$ Astronomy, University of Southampton, Highfield, Southampton SO17 1BJ, UK }
\affiliation{\small   Department of Physics $\&$ Astronomy, Uppsala University, Box 516, SE-751 20 Uppsala, Sweden.}

\begin{abstract}
We establish the possibility of having a pseudo-Nambu-Goldstone boson (pNGB) Dark Matter (DM) candidate in the inspired Froggatt-Nielsen Singlet Model (iFNSM) wherein  a direct connection exists between the DM mass and new flavon symmetry-breaking scale. 
We find a considerable allowed region of parameter space for the ensuing  pseudoscalar DM, which is dependent upon the flavon Vacuum Expectation Value (VEV) and Yukawa couplings, over which it may be possible to explain the fermion mass hierarchy. Finally, we choose a Benchmark Point (BP) and perform detailed collider analyses to probe this DM state in the context of Run 3 of the Large Hadron Collider
(LHC). Specifically, in this model, one obtains large missing transverse energy ($\slashed{E}_T$) when the DM particle is resonantly produced from the decay of a heavy Higgs field, along with multiple jets from  Initial State Radiation (ISR). Thus, the ensuing $\slashed{E}_T$ + $n\, {\rm jets}~(n \geq 1)$ signature is an excellent probe of DM in this construct.
\end{abstract}
 
\maketitle

\section{Introduction}
Explaining the observed hierarchies in fermion masses and mixing continues to be a profound enigma within the Standard Model (SM) framework of particle physics. They are characterized by the Yukawa couplings spanning a vast range of at least six orders of magnitude coupled with the Cabibbo-Kobayashi-Maskawa (CKM) matrix elements exhibiting values that vary from unity down to the minuscule order of $10^{-3}$.
Numerous extensions Beyond the SM (BSM)~\cite{Froggatt:1978nt, Babu:1999me, Giudice:2008uua, Ghosh:2015gpa, Altmannshofer:2015esa, Altmannshofer:2016zrn, Dery:2016fyj, Arroyo-Urena:2022oft, Huitu:2016pwk, Berger:2014gga, Diaz-Cruz:2014pla, Arroyo-Urena:2018mvl, delaVega:2021ugs, Bauer:2016rxs, Greljo:2019xan, Giese:2019khs, Baldes:2016gaf, Feruglio:2015jfa, Qiu:2023igq, Mandal:2023jnv, Abbas:2023ion, Alvarado:2017bax,Abbas:2022zfb, Abbas:2018lga} have been put forth to elucidate these hierarchical patterns.
In a seminal paper~\cite{Froggatt:1978nt}, Froggatt and Nielsen introduced an Abelian flavor symmetry, $U(1)_F$, which helps to explain this Yukawa hierarchy. The main modified Yukawa term of the Lagrangian was $\mathcal{O}=y_q~\left(\frac{S}{\Lambda}\right)^{n}~ \bar Q~ 
H ~ q_{R} +y_\ell~\left(\frac{S}{\Lambda}\right)^{n}~ \bar L~ 
H ~ \ell_{R}  \,$. The symbols $Q$ and $L$ refer to the left-handed quark and lepton doublets, respectively, while $q_R$ and $\ell_R$ represent the right-handed singlet quarks and leptons, respectively. This puzzle can potentially find a resolution through the introduction of new physics parameters, specifically, a new physics scale denoted as $\Lambda$ and the VEV ($v_s$) associated with an additional singlet scalar known as the flavon ($S$), thereby defining the FNSM. The ratio $v_s/\Lambda$ should be less than $1$. Introducing a non-zero value for $v_s$ breaks the flavon symmetry at that particular scale.
It is to be noted that the Froggatt-Nielsen (FN) paradigm, in its original formulation, does not explicitly determine the scale of flavor breaking, $v_s$, nor the new physics scale, $\Lambda$.
The adapted versions of this FN mechanism, as proposed by Babu and Nandi \cite{Babu:1999me} as well as Giudice and Lebedev \cite{Giudice:2008uua}, establish a connection between the flavor breaking scale and the Electro-Weak  Symmetry Breaking (EWSB) scale. They mainly changed $S/\Lambda\rightarrow H^\dagger H/\Lambda^2$ to have this connection.
Phenomenological constraints arising from measurements of the SM Higgs boson mass and signal strengths effectively rule out both the mentioned models. It is noteworthy, though, that more recent alterations, represented by $S/\Lambda\rightarrow H_u^\dagger H_d/\Lambda^2$, within the framework of the 2-Higgs Doublet Model (2HDM), as discussed in \cite{Bauer:2015fxa}, remain a viable avenue of investigation. In this model, $H_u$ and $H_d$ correspond to two scalar doublets coupled to up- and down-type fermionic (charged) particles. However, the introduction of new Higgs signal strength data has significantly narrowed down the viable parameter space in this scenario.

In this study, we introduce a novel implementation of the Froggatt-Nielson mechanism within the SM to elucidate the origin of the observed mass hierarchy and quark-mixing structure. Notably, our approach avoids the necessity of imposing a continuous $U(1)_F$ symmetry. Instead of relying on a continuous Abelian $U(1)_F$ symmetry, we employ straightforward discrete $\mathcal{Z}_4$ symmetry within the SM framework. A similar approach was used in the literature in a different context~\cite{Abbas:2022zfb, Abbas:2018lga}. Here, we focus on a different aspect, keeping the same motivation for the fermion masses. The $\mathcal{Z}_4$ charges of all the particles are shown in table~\ref{tab:charge}. The complex scalar field $S$ has $\mathcal{Z}_4$ charge $``i"\,(i=\sqrt{-1})$. The fermionic combinations ($f_L \,f_R$) transform as $``-1"$, while the SM Higgs doublet transforms as $``+1"$ under $\mathcal{Z}_4$. 
In our current analysis, we modify the conventional FN model by replacing the term $S/\Lambda$ with $\left(S^2+ S^{*2}\right)/\Lambda^2$. This alteration also results in the breaking of the $\mathcal{Z}_4$ symmetry like the original Abelian $U(1)_{F}$ symmetry of the model once the $S$ field acquires a VEV, as $S=\frac {(v_s + S_R + i S_I )}{ \sqrt 2 }$.
It is also to be noted that the imaginary component $S_I$ of $S$ is an NG boson if there is a global $\mathcal{Z}_4$ symmetry in the model. The imaginary component $S_I$ can obtain a mass if the $\mathcal{Z}_4$  symmetry is softly broken, then it becomes a pNGB and its mass can be naturally small compared to the VEV of $S$.  We introduced an additional $\mathcal{Z}_2$ symmetry to ensure that the singlet scalar field transforms as $S\to S^*$, which forbids the $\left(S^2 - S^{*2}\right)/\Lambda^2$ term in the Lagrangian~\cite{Kannike:2019mzk, Gross:2017dan}. All other particles are even under $\mathcal{Z}_2$ symmetry. Notably, this new $\mathcal{Z}_2$ and $\mathcal{Z}_4$ symmetries instead of old $U(1)_F$ imposes a constraint that prevents the decay of the imaginary component of the complex singlet scalar field $(S_I)$. Consequently, such a pNGB becomes a stable particle and, contingent upon the specific parameter space considered, emerges as a promising candidate for DM.
One of the main and important parts of this paper is that we also establish a connection between the (pNGB) DM mass with the above flavor-breaking VEV and the new physics scale via $\left(S^2+ S^{*2}\right)/\Lambda^2$ in this construction.
To the best of our knowledge, this particular connection within the framework of the inspired FN scenario has not been previously explored or discussed in existing literature.
\begin{table}[htb!]
\centering
\begin{tabular}{|c|c|c|}
\hline
~~~~~~~ Particles~~~~~~~ &~~~~~~~ $\mathcal{Z}_4$ Charges ~~~~~~~ \\
\hline
$S$&    $i=\sqrt{-1}$\\  
$S^*$&    $-i$\\  
$\Phi$&    $1$\\
$Q_{i,L},\,\,\psi_{i,L}$ &  $1$\\
$d_{i,R},\,\,\ell_{i,R}$ &  $-1$\\
\hline
\end{tabular}
\caption{ \it The $\mathcal{Z}_4$ charges of all the particles. The fields $Q_L=(u, \,\,\, d)_L^T$ and $\psi_L=(\nu_\ell,\,\,\,\ell)_L^T$ are the SM left-handed fermions whereas $d_R,u_R$ and $\ell_R$ are the SM right-handed fermions. All other particles are even under $\mathcal{Z}_2$ symmetry, except the singlet scalar field, which transforms as $S\to S^*$.}
\label{tab:charge}
\end{table}

The DM particle within this framework is constrained by a plethora of experimental data, primarily from coupling measurements of the SM-like Higgs boson discovered in 2012 and several DM particle searches. The presence of DM particles at colliders is usually probed through signatures like mono-jet plus large $\slashed{E}_T$ \cite{Fox:2011pm,Abdallah:2015ter,Felix,ATLAS:2021kxv, CMS:2021far,ATLAS:2020uiq,CMS:2018ffd,Ghosh:2021noq, Ghosh:2022rta, Ghosh:2023xhs}. The reason is that we cannot directly observe invisible particles (i.e., trigger on nothing). Therefore, we need extra visible ones to trigger on. The mono-jet signal is rather copious, owing to the predominant QCD interactions at the LHC. Such a signature is mostly expected to originate from QCD ISR, and demanding the final state jet to have a large transverse momentum helps to reduce the dominant SM backgrounds. Mono-photon is also a channel pursued at the LHC but has less sensitivity in comparison due to a smaller cross-section~\cite{CMS:2017qyo}.
Furthermore, both the CMS and ATLAS collaborations at the LHC have also searched for DM particles produced in association with a $Z$ (mono-$Z$) or a Higgs boson (mono-Higgs), when these bosons decay hadronically or leptonically (or even to a pair of photons for the Higgs boson) \cite{CMS:2020ulv,ATLAS:2021shl,ATLAS:2021jbf,ATLAS:2021gcn}. Mono-$W$ is also a search channel~\cite{ATLAS:2014pna,ATLAS:2018nda}. Finally,  searches for DM particles produced from the decay of exotic scalars in events with Vector-Boson Fusion (VBF) signatures are also studied at the LHC \cite{CMS:2022qva,ATLAS:2022yvh}. 

 In this work, we study the detection prospects for parameter space points that satisfy the DM relic density constraints and comply with DM direct and indirect detection limits. Within the FNSM model, the primary mechanism for producing the pNGB DM particle (a pseudoscalar denoted as $A_F$) at the LHC involves the production of $H_F$ (the scalar counterpart, known as the flavon), which subsequently decays into a pair of $A_F$ particles. Additionally, QCD-induced initial-state radiation (ISR), potentially resulting in one or more jets, may also be present along with these new invisible particles.
Similarly, one would also expect QED-induced ISR photon, leading to mono/di-photon plus missing $E_T$ 
signal. However, it can be ignored compared to the jets plus missing $E_T$ signal~\cite{CMS:2017qyo}. As for mono-$W^\pm$,  the emerging lepton plus $\slashed{E}_T$ final state does not offer a good statistical significance for discovering these DM particles, while hadronic $W$ decays have significant background in comparison~\cite{ATLAS:2014pna,ATLAS:2018nda}. The alternative probes through VBF production modes or associated production with a $Z$ or Higgs boson are also found to be less significant for our forthcoming BPs. Therefore, we focus on the monojet + $\slashed{E}_T$ final state topology, where we demand (at least) one highly energetic jet ($p_T >$  150 GeV) along with large $\slashed{E}_T$ in the context of the 14 TeV run of the LHC with integrated luminosity 300 ${\rm fb}^{−1}$ (Run 3). In this connection, as mentioned, we work with BPs in order to facilitate searches for mono-jet events, including $\slashed{E}_T$.

The paper is structured as follows. Section~\ref{sec:2} provides a concise overview of the model framework. Section~\ref{sec:3} is dedicated to discussing the relevant theoretical and experimental constraints applicable to the model.
Moving on to Section~\ref{sec:4}, we present the setup of our numerical analysis. Section~\ref{se:col_an} is focused on extracting the signature of the DM signal arising from the decay of the flavon. Finally, Section~\ref{sec:5} summarizes the key findings and draws conclusions based on our study.

\section{The Model}
\label{sec:2}
We now focus on some relevant theoretical aspects of what we will refer to as the inspired FNSM.

\subsection{The (Pseudo)Scalar Sector} 
The scalar sector of this model includes the SM Higgs doublet denoted as $\Phi$ and a single $SU(2)_L$ singlet complex FN spin-0 field referred to as the flavon field, $S$. We can express these fields using the following parameterization: 
\begin{eqnarray} 
	& \Phi = \left( \begin{array}{  c} 0 \\ \frac{  v + \phi^0}{\sqrt 2}\\
	\end{array}  \right), \label{dec_doublets}&\\ 
	& S = \frac {(v_s + S_R + i S_I )}{ \sqrt 2 }   , \label{dec_Singlet} &
\end{eqnarray}
where $S_R$ is the  scalar component of $S$ and $S_I$ is the pseudoscalar component of $S$. 
The notations $v$ and $v_s$ represent the VEVs of the SM Higgs doublet and FN  singlet $S_R$, respectively.
To ensure that the scalar potential remains invariant under the $\mathcal{Z}_2$ and $\mathcal{Z}_4$ symmetries, the Standard Model (SM) Higgs doublet is left unchanged under both symmetries. In contrast, the FN singlet  transforms as $S \to S^*$ under $\mathcal{Z}_2$ and as $S \to i\, S$ under $\mathcal{Z}_4$, as summarized in Table~\ref{tab:charge}.
\\
In general, a scalar potential (if allowed by any additional symmetry) allows the possibility of a complex VEV, $\langle S\rangle_0=\frac{v_s}{\sqrt{2}}e^ {i\xi} $. We focus on the scenario where the Higgs potential is CP-conserving. The phase parameter $\xi$ to zero due to the symmetries $\mathcal{Z}_2$ and $\mathcal{Z}_4$.
In the context of CP-conservation and considering the introduced $\mathcal{Z}_2$ and $\mathcal{Z}_4$ symmetries, the Higgs potential can be expressed as follows:
\begin{eqnarray} \label{potential} 
	V_0=-\frac {1}{2} m_1^2\Phi^ \dagger \Phi-\frac{1}{2} m_{2}^2 
S^*S +\frac {1}{2} \lambda_1 \left(\Phi^ \dagger \Phi\right)^2+\lambda_2
\left(S^*S\right)^2 +\lambda_ {3} \left(\Phi^ \dagger \Phi\right)\left(S^* S\right). 
\end{eqnarray}
The $\mathcal{Z}_4$ symmetry of this scalar potential is spontaneously broken by the VEVs of the spin-0 fields $S$ (keeping $\mathcal{Z}_2$ intact), resulting in the emergence of a massless Goldstone boson in the physical spectrum. To impart mass to this Goldstone boson, we introduce the following soft $\mathcal{Z}_4$ breaking term (while still prohibiting the decay of $S_I$) to the potential:
\begin{eqnarray}
V_{\rm soft} = -\frac{m_3^2}{2} \left (S^{2} + S^{*2} \right).  
\end{eqnarray}
 After $\mathcal{Z}_4$ breaking the massless Goldstone boson $S_I$  gets a mass and becomes a pNGB depending on $m_3$. The $\mathcal{Z}_2$ remains unbroken, which helps to stabilize the imaginary part of the complex singlet scalar. The complete scalar potential is given by
\begin{eqnarray}
V = V_0 + V_{\rm soft}. 
\end{eqnarray}

The presence of the $\lambda_ {3} \left(\Phi^ \dagger \Phi\right)\left(S^* S\right)$ term in the scalar potential allows for the mixing between the Higgs and flavon fields, which occurs after both the $\mathcal{Z}_4$ and EWSB. This mixing contributes to the mass parameters for both the Higgs and flavon fields, as demonstrated below. The soft $\mathcal{Z}_4$ symmetry breaking term, denoted as $V_{\rm soft}$, is responsible for generating the mass of the pseudoscalar (CP-odd) flavon field $S_I$. We now derive the following relationships between the parameters of the potential after applying the minimization conditions:
\begin{eqnarray} 
	m_{1} ^2  &=&  v^2 \lambda_1 + v_s^2 \lambda_{3},   \\
	m_{2} ^2 &=& -2 m^2_{3} + 2 v_s^2 \lambda_2 + v^2 \lambda_{3}.
\end{eqnarray}
Since all the parameters of the scalar potential are real, there is no mixing between the real and imaginary parts of the potential. As a result, the CP-even mass matrix can be expressed in the $(\phi_0, S_R)$ basis as:
\begin{equation} 
	M^2_S =
	\left( \begin{array}{cc} 
		\lambda_1 v^2      &  \lambda_{3} v v_s \\
		\lambda_{3}v v_s   &  2 \lambda_2 v_s^2
	\end{array}  \right).
\end{equation} 
The mass eigenstates are determined through the standard $2\times 2$ rotation and can be expressed as follows:
\begin{eqnarray} 
	\phi^0   &=& \  \  \  \  \cos \  \alpha \  h + \sin \  \alpha  \  H_F,   \\
	S_R    &=& -\sin \  \alpha \  h + \cos \  \alpha \  H_F.
\end{eqnarray}
Here, $\alpha$ stands for the mixing angle. In this context, we identify the particle denoted by $h$ as the SM-like Higgs boson with a mass of $M_h = 125.5$ GeV, while the mass eigenstate $H_F$ corresponds to the CP-even flavon.
The CP-odd flavon, denoted as $A_F\equiv S_I$, possesses a mass given by $M^2_{A_F} = 2m_3^2$. It is important to note that both $H_F$ and $A_F$ are regarded as having greater masses than the SM-like Higgs boson $h$.
In this model, our analysis will involve the utilization of the mixing angle $\alpha$ along with the physical masses $M_h$, $M_{H_F}$ and $M_{A_F}$. These physical masses are interconnected with the quartic couplings of the scalar potential in Eq.~(\ref{potential}), as: 
\begin{eqnarray}\label{eq:relate}
\lambda_1&=& \frac{ \cos\alpha^2 M_h^2+\sin\alpha^2 M_{{H_F}}^2}{v^2},\nn\\
\lambda_2&=& \frac{M_{{A_F}}^2+{\cos\alpha }^2 M_{{H_F}}^2+{\sin\alpha }^2 M_h^2}{2 v_s^2},\\
\lambda_3&=& \frac{ \cos\alpha \, \sin\alpha }{ v v_s} \, ( M_{{H_F}}^2 -  M_h^2).\nn
\end{eqnarray} 
Our analysis involves treating the mixing angle $\alpha$, the VEV  of the FN singlet $v_s$ and the masses of its (pseudo)scalar field $M_{H_F}$($M_{A_F}$) as free parameters. Depending on specific input values, this CP-odd flavon, i.e., pNGB $A_F$ can  serve as a viable DM candidate.
The coupling strength of this particle to the CP-even Higgses are: 
    \begin{equation}
    g_{A_F A_F h}=-i(\cos\alpha \, \lambda_{3} v_{\rm SM} - 2 \sin\alpha\,  \lambda_{2} v_s )=i\,\sin\alpha\,(M_h^2+M_{{A_F}}^2)/v_s
    \end{equation}
    and 
    \begin{equation}
     g_{A_F A_F H_F}=-i(\sin\alpha \, \lambda_{3} v_{\rm SM} + 2 \cos\alpha \, \lambda_{2} v_s)=i\,\cos\alpha\,(M_{{H_F}}^2+M_{{A_F}}^2)/v_s,   
    \end{equation} 
    respectively.
\subsection{The Yukawa Sector} 
The new model, i.e., the iFNSM, in addition to the new complex scalar singlet,  also invokes the FN mechanism \cite{Froggatt:1978nt} with a discrete $\mathcal{Z}_2$ and $\mathcal{Z}_4$ symmetries instead of an continuous abelian $U(1)_F$ symmetry.
We can get the Yukawa part $Y^d_{ ij }\, \overline{Q_{i,L}}\, d_{j,R}   \Phi 
	+ Y^u_{ ij } \, \overline{Q_{i,L}} \, u_{j,R} \tilde \Phi+ Y^\ell_{ ij }  \, \overline{\psi_{i,L}} \,\ell_{j,R} \Phi  + \rm H.c$ of the SM Lagrangian from the point of
view of an Effective Field Theory (EFT) invariant under $\mathcal{Z}_4$ transformation that distinguishes amongst fermion families.  All the particles are invariant under the $\mathcal{Z}_2$ transformation, except for the imaginary part of the complex scalar field. The re-phasing of the SM fields under the $\mathcal{Z}_4$ transformation implies $Q_i \rightarrow  Q_i$, $d_i \rightarrow - d_i$, $u_i \rightarrow - u_i$, $\psi_i \rightarrow  \psi_i$ and $\ell_i \rightarrow - \ell_i$. Here, $Q_L=(u, \,\,\, d)_L^T$ and $\psi_L=(\nu_\ell,\,\,\,\ell)_L^T$ are the SM left-handed fermions whereas $d_R,u_R$ and $\ell_R$ are the SM right-handed fermions. Furthermore, the indices $i,j$ run over all three generations. Finally, $\tilde \Phi = i \, \tau_2\, \Phi $, where $\tau_2$ stands for the second Pauli  matrix.
\\
 The effective iFNSM $\mathcal{Z}_2$ and $\mathcal{Z}_4$-invariant Lagrangian can be written as:
\begin{align} 
	\mathcal{L}_ Y &=    \rho^d_{ ij } \frac{\left(S^2+ S^{*2}\right)}{\Lambda^2} \,  \overline{Q_{i,L}} \, d_{j,R}   \Phi 
	+ \rho^u_{ ij }  \frac{\left(S^2+ S^{*2}\right)}{\Lambda^2}\,  \overline{Q_{i,L}}\, u_{j,R} \tilde \Phi
	\nonumber\\&+ \rho^\ell_{ ij }  \frac{\left(S^2+ S^{*2}\right)}{\Lambda^2}\,  \overline{\psi_{i,L}}\, \ell_{j,R} \Phi  + \rm H.c., 
	\label{eq:yuk1} 
\end{align} 
One has: 
\begin{align} 
	\frac{\left(S^2+ S^{*2}\right)}{\Lambda^2}  &= \left( \frac{v_s^2 + 2v_s \, s + s^2 + A_F^2}{   \Lambda^2}  \right) \nonumber\\
	&= \left( \frac{v_s^2}{ \Lambda^2}  \right) \, \left( 1 + (\frac{2 v_s \, s + s^2 + A_F^2}{v_s^2}) \right).
\end{align} 
This leads to the following fermion couplings, after replacing for the mass eigenstates in $\mathcal{L}_ Y $:
\begin{align} 
	\mathcal{L}_ Y &= \frac{1}{\sqrt{2}}\, (\overline{d_{i,L}}  \, y^d_{ij}  \, d_{j,R}+\overline{u_{i,L}} \,  y^u_{ij}  \, u_{j,R}+\overline{\ell_{i,L}}  \, y^\ell_{ij}  \, \ell_{j,R})\, (v_{\rm SM}+\phi^0)+ \frac{1}{\sqrt{2}}\, (n_{Q_id_j}  \, \overline{d_{i,L}}  \, y^d_{ij}  \, d_{j,R} \nonumber\\
	&+n_{Q_iu_j} \,  \overline{u_{i,L}} \,  y^u_{ij}  \, u_{j,R} +n_{L_i\ell_j}  \, \overline{\ell_{i,L}}  \, y^\ell_{ij}  \, \ell_{j,R})\, (v_{\rm SM}+\phi^0)\, (2 v_s \, s + s^2 + A_F^2).
\end{align} 
Here, $y^f_{ij}= \left( \frac{v_s^2}{ \Lambda^2}  \right) \, \rho^f_{ ij }$ with $\phi^0=\cos \  \alpha \  h + \sin \  \alpha  \  H_F  $ and $s =-\sin \  \alpha \  h + \cos \  \alpha \  H_F $.
The mass and interaction terms describing the coupling of fermions with scalar particles are as follows:
\begin{align} 
	\mathcal{L}_ {Y} &= \frac{v_{\rm SM}}{\sqrt{2}}\, (\overline{Q_L} \, \hat{y}^d d_R+\overline{Q_L}\,\hat{y}^u u_R+\overline{\psi_L} \,\hat{y}^l \ell_R)+\mathcal{L}_ {\rm int}.
\end{align} 
The Yukawa couplings $\hat{y}^f$ are real and diagonal  matrices, whose
eigenvalues are related to the charged fermion masses $M_f$ via $\hat{y}^f=\sqrt{2} M_f/v_{\rm SM}=V^{f\dagger}_L y^f_{ij} V^f_R$, with $V$ being the diagonalizing matrix. Then,
\begin{align} 
	\mathcal{L}_ {\rm int} &=  (\cos\alpha \, \frac{M_f}{v_{\rm SM}} -2\sin\alpha \,Z_{ij}\, \frac{ v_{\rm SM}}{\sqrt{2} \,v_s} ) \,h\bar{F}\,F + (\sin\alpha \, \frac{M_f}{v_{\rm SM}} +2\cos\alpha \,Z_{ij}\,  \frac{ v_{\rm SM}}{\sqrt{2} \,v_s} ) \,H_F \bar{F}\,F \nn\\
\label{eq:fermiYukfinal}
	&+ \frac{\sin\alpha \, (\sin\alpha \, v_{\rm SM} -2 \cos\alpha \, v_s) }{\sqrt{2} v_s^2} Z_{ij} \,h^2\bar{F}\,F + \frac{\cos\alpha \, (\cos\alpha \, v_{\rm SM} +2 \sin\alpha \, v_s) }{\sqrt{2} v_s^2} Z_{ij} \,H_F^2\bar{F}\,F\\
	& + \frac{(2\,\cos2\alpha \, v_s - \sin2\alpha \, v_{\rm SM}) }{\sqrt{2} v_s^2} Z_{ij} \,hH_F\bar{F}\,F + \frac{v_{\rm SM}}{\sqrt{2} v_s^2} Z_{ij} \,A_F^2\bar{F}\,F,\nn
\end{align} 
where $Z_{ij}= V^{f\dagger}_L  \,y^f_{ij} V^f_R =  V^{f\dagger}_L  \,\left( \frac{v_s^2}{  \Lambda^2}  \right) \, \rho^f_{ ij }\, V^f_R$ and $F$ stands for the charged fermions in the mass basis. The above Lagrangian implies that the primary factors influencing the DM density in this model are the DM mass, the VEVs of the singlet (pseudo)scalars, the parameters denoted as $n_{ij}$ and the Yukawa couplings.

\section{Constraints}\label{sec:3}

In order to perform a rigorous numerical analysis of pNGB DM within the framework of the iFNSM model, it is imperative to constrain several key parameters carefully. These parameters encompass: $(i)$ the mixing angle $\alpha$ between the real components of the doublet field $\Phi$ and the FN singlet field $S$; $(ii)$ the FN singlet VEV ($v_s$); $(iii)$ the masses of the heavy scalar and pseudo-scalar fields $M_{H_F}$ and $M_{A_F}$, respectively; $(iv)$ the elements of the ${Z}_{ij}$ matrix, which represent the mixing of the various SM fermionic fields within the iFNSM. Their precise determination is crucial as they profoundly impact the iFNSM  particle spectrum, interactions and, ultimately, its DM properties.

These parameters undergo stringent constraints imposed by a combination of theoretical considerations, such as ensuring absolute vacuum stability, maintaining perturbativity and preserving the unitarity of scattering matrices. Additionally, they are subject to further scrutiny based on a plethora of experimental data, primarily focusing on the findings from the measurements of various couplings of the SM-like Higgs boson discovered in 2012 and preformed at the LHC. Then, Lepton Flavor Violating (LFV) processes (like $\tau\to 3\mu$, $\mu\to 3e$, $\tau\to \mu\gamma$, $\mu\to\ e\gamma$), $B_s^0\to\mu^+\mu^-$ along with the total decay width of the Higgs boson receive large modification in the presence of these new Yukawa couplings. 
All such constraints are implemented in our analysis of the iFNSM scenario.

\begin{figure}[h!]
	\begin{center}
		\subfigure[]{
			\includegraphics[scale=0.340]{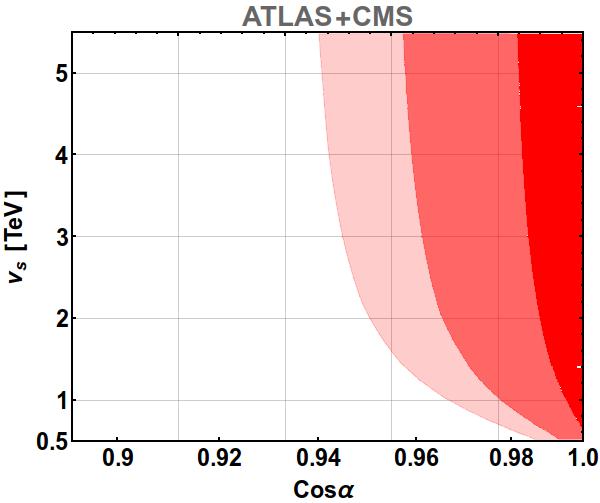}}	
					\subfigure[]{
			\includegraphics[scale=0.350]{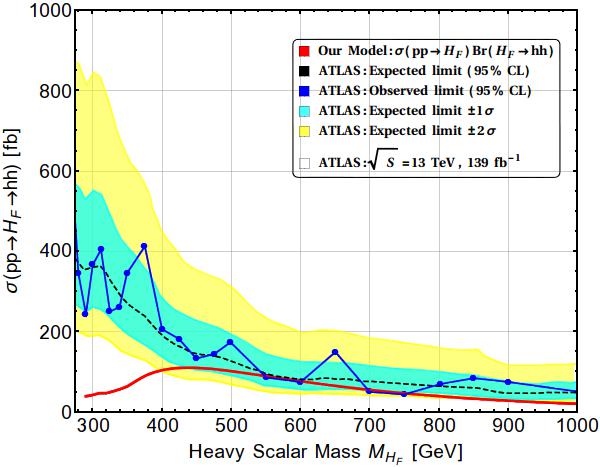}}
		\caption{ \it (left) Allowed $1\sigma$ (dark red), $2\sigma$ (medium  red) and $3\sigma$ (light red) regions, for a global $\chi^2-$fit to ATLAS and CMS data from measurements of the SM-like Higgs boson decays. (Right) Observed and expected upper limits on the cross-section for di-Higgs production (compare to  Ref.~\cite{ATLAS:2021ifb}) through $H_F$ as a function of the particle mass $M_{H_F}$ as obtained through the process $pp\to H_F\to hh~(h,\to b\bar b, h\to \gamma \gamma)$.}
		\label{fig:DarkPlotkinmix}
	\end{center}
\end{figure}

 Ref.~\cite{Arroyo-Urena:2022oft} extensively explores how various parameters and predictions within the FNSM model are subjected to rigorous scrutiny and alignment with theoretical requirements and experimental data. 
The fermion sector of our model is quite different from that of Ref.~\cite{Arroyo-Urena:2022oft}, though, hence, the constraints are also different here. However, the stability, perturbativity, and unitarity criteria remain similar as we have the same (pseudo)scalar and vector sectors. In Figure~\ref{fig:DarkPlotkinmix}(left), we present the allowed parameters in the ($v_s,\cos\alpha$) plane from the LHC Higgs signal strength measurements~\cite{Cepeda:2019klc,ATLAS:2020qdt}. We have then done a $\chi^2-$fit to get the allowed parameters over such a plane. Here, the heavy (pseudo)scalar field masses do not play any major role in this global $\chi^2-$fit plot as we have chosen $\Lambda=100$ TeV. We also present  the observed and expected upper limits on the di-Higgs production cross-section in Figure.~\ref{fig:DarkPlotkinmix}(right) for a singlet scalar VEV $v_s=1$ TeV (as in Ref.~\cite{ATLAS:2021ifb}). 

Di-Higgs production and decay, as discussed in Ref.  \cite{ATLAS:2021ifb}, eventually featuring final states composed of pairs of bottom quarks and photons, denoted as $pp\to H_F\to hh$ with one $h$ decaying to $b\bar{b}$ and the other via $h\to \gamma \gamma$, imposes a highly stringent upper bound on the mass of the heavy CP-even Higgs bosons $M_{H_F}$. In this analysis we safely use $M_{H_F}>250$ GeV.
The LFV processes put bounds on the matrix element $Z_{ij}$, flavon VEV $v_s$ and flavon mass $M_{H_F}$. The decay width for the process $\tau\to \mu \gamma$ can be written as~\cite{Harnik:2012pb}:
\begin{eqnarray}
\Gamma[\tau \to \mu \gamma] &\approx  \frac{\alpha m_\tau^5 v_{\rm SM}^4}{4608\, \pi^4 M_{H_F}^4 v_s^4} \,\Big| Z_{\tau\tau} \,Z_{\mu\tau} \left[ 3 \, \log(M_{H_F}^2/m_\tau^2)-4\right]\Big|^2.
\label{eq:lfv1}
\end{eqnarray}
The expression for  $\Gamma[\mu\to e \gamma]$ can be obtained by simply replacing $\tau$ to $\mu$ and $\mu$ to $e$ in Eq.~(\ref{eq:lfv1}). Similarly the expression for $\tau \to 3 \mu$ and $\mu \to 3 e$ can be written using Ref.~\cite{Harnik:2012pb}.
Certainly, in the context of the iFNSM, we can assume that the coefficients $Z_{ij}$ are of order $\mathcal{O}(1)$ to circumvent the stringent LFV constraints, as discussed in Refs.~\cite{Baldini:2018nnn, Workman:2022ynf}.

\begin{figure}[h!]
	\begin{center}
{
			\includegraphics[scale=0.3500]{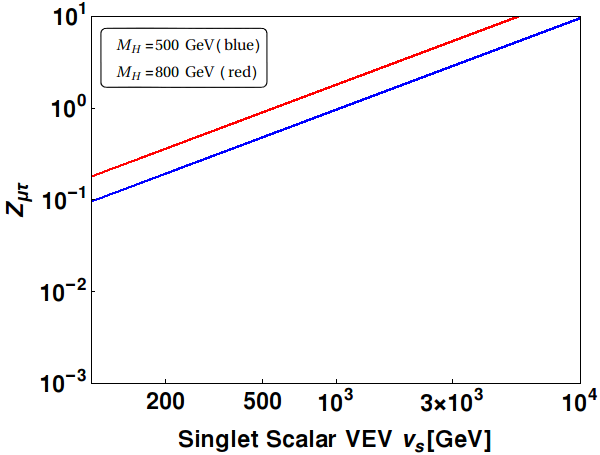}}	
		\caption{ \it Bounds coming from the recent FNAL result for $(g-2)_\mu$ \cite{Muong-2:2023cdq}, when combined with the earlier results from BNL+FNAL \cite{Muong-2:2006rrc,Muong-2:2021ojo}, giving $\delta a_\mu = (2.49\pm 0.48)\times 10^{-9}$. The regions above the lines are ruled out. }
		\label{fig:mug2}
	\end{center}
\end{figure}

The most recent measurement of the muon anomalous magnetic moment, $(g-2)_\mu$, at FNAL~\cite{Muong-2:2023cdq}, when combined with the earlier results from BNL+FNAL \cite{Muong-2:2006rrc,Muong-2:2021ojo}, yielding $\delta a_\mu = (2.49\pm 0.48)\times 10^{-9}$, imposes stringent constraints on the model parameters. The largest and dominant contribution to $\delta a_\mu$ in this scenario is given by ~\cite{Harnik:2012pb}:
\begin{equation}
\delta a_\mu \approx \frac{m_\mu  m_\tau }{16\, \pi^2 M_{H_F}^2} \, Re[Z_{\mu\tau} Z_{\tau\mu}] \left[ 2 \, \log(M_{H_F}^2/m_\tau^2)-3\right].
\end{equation}
The resulting allowed region in the ($v_s, {Z}_{\mu\tau}$)  
plane is illustrated in Figure~\ref{fig:mug2}. It is worth noting here that the behavior of the ${Z}{\mu\tau}$ matrix element exhibits an increasing trend as $v_s$ increases and a decreasing trend as $v_s$ decreases.
This behavior aligns with expectations since the ${H_F \mu\tau}$ coupling is influenced by ${Z}{\mu\tau}/v_s$. To ensure a reliable assessment of the observables under DM investigation, we opt for conservative values when determining ${Z}_{\mu\tau}$ and $v_s$.

In conjunction with other cosmological measurements, the Wilkinson Microwave Aniso\-tropy Probe (WMAP) satellite has yielded valuable constraints on the relic density of DM. The value for the DM relic density is determined to be $\Omega h^2=0.1198\pm 0.0012$~\cite{Aghanim:2018eyx}.
In recent times, direct detection experiments, such as Xenon-1T~\cite{Aprile:2018dbl}, LUX-ZEPLIN (LZ)~\cite{LZ:2022ufs} and PANDAX~\cite{PandaX:2016pdl}  have been at the forefront of the search for Weakly Interacting Massive Particles (WIMPs) as potential candidates for DM. 
In this model, we designate the DM candidate as a scalar, and consequently, the relevant constraint stems from the Spin-Independent (SI) cross-section with nucleons.
One can remove the tree-level direct detection cross-section by adjusting the scalar potential parameter~\cite{Alanne:2020jwx, Azevedo:2018exj}; however, at one-loop, it could again give a non-zero direct detection cross-section, which may put constraints on the model parameters. This adjustment does not apply to our model. We present such details regarding direct detection in Appendix~\ref{eq:ddcross}.
For DM masses spanning the range $\mathcal{O}(10-1000)$ GeV, the most stringent constraints are derived from the Xenon-1T experiment~\cite{Aprile:2018dbl}, where the cross-section values are typically around $\mathcal{O}(10^{-46}-10^{-45})~{\rm cm}^2$.
The null results from these experiments, along with data from collider experiments, which encompass measurements related to the invisible widths of the SM Higgs boson and gauge bosons, have imposed constraints on both the couplings and masses of the DM particles.
In the context of a singlet pseudoscalar DM model, it becomes feasible to account for the findings from various indirect DM detection experiments~\cite{Cline:2013gha, Cheung:2012xb} within certain parameter regions. However, our focus does not involve a deep dive into these specifics. Estimations of this nature require a thorough grasp of astrophysical backgrounds and necessitate assumptions about the DM halo profile, which inherently introduces some degree of arbitrariness.
For a comprehensive overview of constraints placed on a singlet pseudoscalar DM model from astrophysical perspectives, one can refer to, for instance, the reviews provided in Refs.~\cite{Cline:2013gha, Cheung:2012xb}. However, we do examine the constraints on 
the  thermal averaged DM annihilation cross-section, particularly for $<\sigma v>{(A_F A_F \rightarrow b\bar{b})}$ and $<\sigma v>{(A_F A_F \rightarrow \tau^+ \tau^-)}$ are consistent ($<\sigma v> < \mathcal{O}(10^{-28})~{\rm cm}^3/{\rm s}$) 
with data from the Fermi Gamma-ray Space Telescope~\cite{Geringer-Sameth:2011wse, Fermi-LAT:2011vow} across the parameter space studied in this analysis. As intimated, 
the iFNSM introduces a flavor symmetry that addresses the mass hierarchy observed in  SM fermions. This hierarchy is directly linked to the DM density, so incorporating these flavor symmetries allows us to explore a broader region of parameter space for the DM.

In this paper, our approach begins by utilizing {\tt FeynRules}~\cite{Alloul:2013bka} to construct the model files. Subsequently, we employ {\tt micrOMEGAs}~\cite{Belanger:2018mqt} to calculate the relic density of the pseudoscalar DM. Furthermore, we have conducted verification of the outcome through the use of {\tt SARAH}~\cite{Staub:2013tta}, which includes the {\tt SPheno}~\cite{Porod:2011nf} mass spectrum calculation integrated into {\tt micrOMEGAs}. In the forthcoming numerical analysis section, we extensively discuss this DM scenario.
\section{DM Analysis }
\label{sec:4}
Here, the complex part of the singlet scalar $A_F$ could not decay due to imposed $\mathcal{Z}_2$ symmetry and has the potential to account for the observed relic density in the Universe via the freeze-out mechanism only. The applicability of the Freeze-in mechanism in this scenario primarily arises from the presence of significant Yukawa couplings within the fermionic sector, directly linked to the masses of fermions (charged leptons and quarks).
We also assume that the DM could undergo annihilation in the early Universe when it was in thermal equilibrium with other particles. As the universe temperature drops below the DM mass, DM freezes out.

\begin{figure}[h!]
	\begin{center}
		\subfigure[]{\includegraphics[scale=0.500]{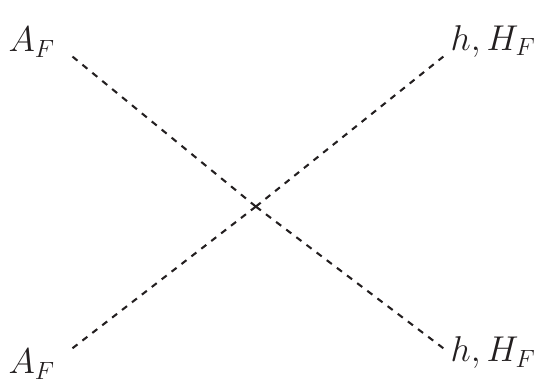}}	
		\subfigure[]{\includegraphics[scale=0.500]{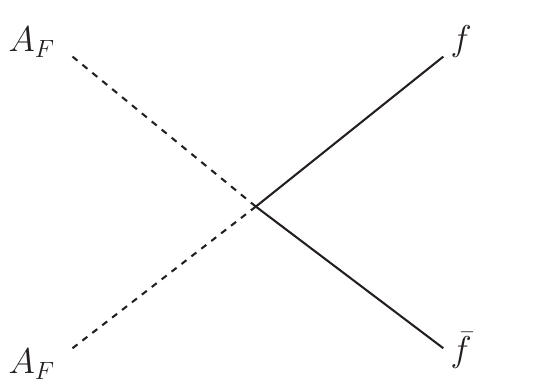}}
		\subfigure[]{\includegraphics[scale=0.500]{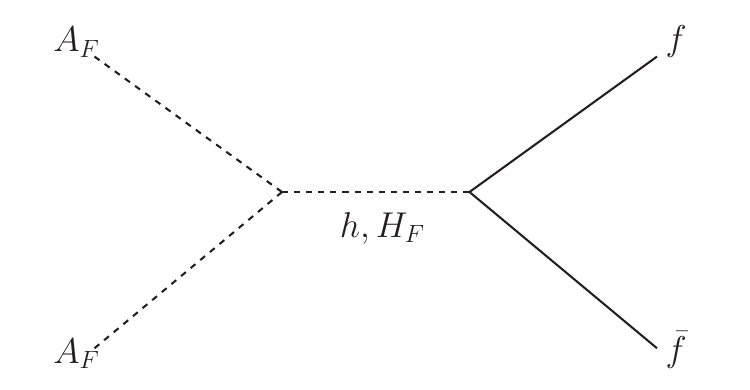}}	
		\subfigure[]{\includegraphics[scale=0.500]{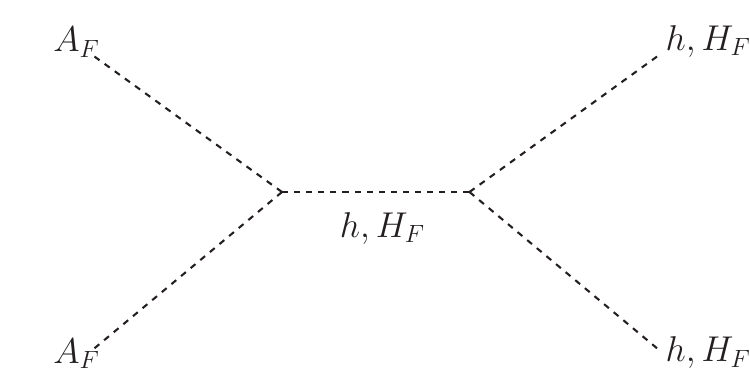}}	
		\subfigure[]{\includegraphics[scale=0.500]{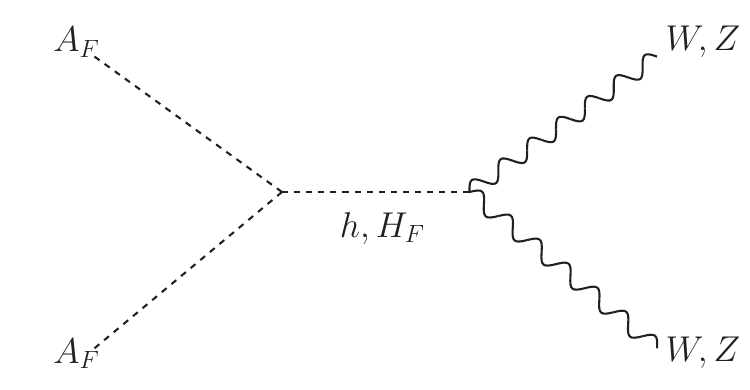}}			%
		\caption{ \it All the possible DM annihilation diagrams into other particles.}
		\label{fig:DMannidiag}
	\end{center}
\end{figure}
The annihilation processes are shown in the Figure~\ref{fig:DMannidiag}. The relic density is influenced by several factors, including the new quartic $\lambda$'s and Yukawa couplings $\rho's$, the VEV of the singlet scalar $v_s$ and the masses of the initial and final state particles involved in the annihilation process. 
In the low-mass range of DM with $M_{A_F}\equiv M_{\rm DM}< 70$ GeV, the dominant contribution to the relic density stems from the $A_F A_F \rightarrow b \bar{b}$ annihilation channel. However, it's worth noting that a substantial portion of this low DM mass region, which could produce the exact relic density within $\Omega h^2=0.1198\pm 0.0012$, are ruled out either from Higgs decay width and/or the direct detection data, except very narrow regions when the  DM mass is close to $M_{h}/2\approx 62.75$ GeV and $M_{A_F}/2$.
\begin{table*}[htb!]
\begin{center}\scalebox{0.8}{
	\begin{tabular}{|p{2.0cm}|p{1.25cm}|p{1.2cm}|p{1.2cm}|p{1.2cm}|p{1.4cm}|p{2.5cm}|p{4.6cm}|}
		\hline
		\hline
		Sl. No. &Mixing $\cos\alpha$& $M_{\rm DM}$ (GeV) & $M_{H_F}$ (GeV) & $v_s$ (TeV)& $\Omega_{\rm DM}h^2$&DD cross-sec (${\rm cm^3/s}$)&~~~~Contributions \\
		\hline
	    \hline
		&&&&&&&$\sigma(A_F \,A_F\rightarrow b \overline{b})~~76\%$\\
		BP-1a	&~0.995&~54.75&~500&1.0&0.1196&$5.83 \times 10^{-45}$& $\sigma(A_F \,A_F\rightarrow q \overline{q})\quad 6 \%$\\
		 &&&&&&&$\sigma(A_F \,A_F\rightarrow l \overline{l})~~4\%$\\
		&&&&&&&$\sigma(A_F \,A_F\rightarrow W W^* )~~13\%$\\
		&&&&&&&$\sigma(A_F \,H^\pm\rightarrow ZZ^*)~~1\%$\\
		\hline
	    \hline
		&&&&&&&$\sigma(A_F \,A_F\rightarrow b \overline{b})~~76\%$\\
		BP-2a	&~0.995&~58.25&~500&5.0&0.1193&$2.86 \times 10^{-48}$& $\sigma(A_F \,A_F\rightarrow q \overline{q})\quad 6 \%$\\
		 &&&&&&&$\sigma(A_F \,A_F\rightarrow l \overline{l})~~4\%$\\
		&&&&&&&$\sigma(A_F \,A_F\rightarrow W W^* )~~12\%$\\
		&&&&&&&$\sigma(A_F \,H^\pm\rightarrow ZZ^*)~~2\%$\\
		\hline
	    \hline
		&&&&&&&$\sigma(A_F \,A_F\rightarrow b \overline{b})~~76\%$\\
		BP-3a	&~0.995&~59.75&~500&10.0&0.1235&$4.6 \times 10^{-49}$& $\sigma(A_F \,A_F\rightarrow q \overline{q})\quad 6 \%$\\
		 &&&&&&&$\sigma(A_F \,A_F\rightarrow l \overline{l})~~4\%$\\
		&&&&&&&$\sigma(A_F \,A_F\rightarrow W W^* )~~12\%$\\
		&&&&&&&$\sigma(A_F \,H^\pm\rightarrow ZZ^*)~~2\%$\\
		\hline
		\hline
	\end{tabular}}
\end{center}
	\caption{\it DM annihilation contributions in the low mass region for a mixing angle $\cos\alpha=0.995$.}
	\label{tabDM:anni1}
\end{table*}
\begin{figure}[htb!]
	\begin{center}
		\subfigure[]{
			\includegraphics[scale=0.3500]{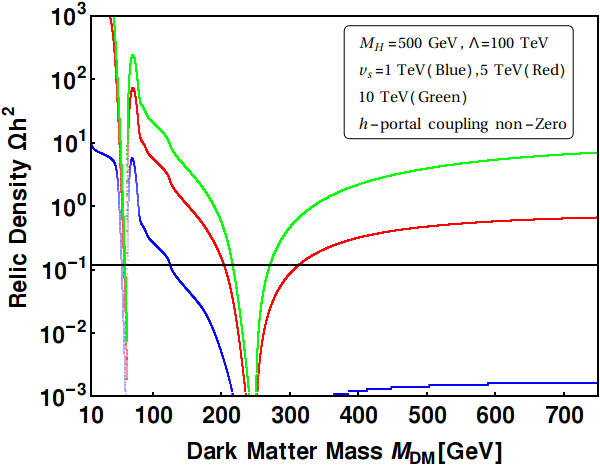}}	
		\subfigure[]{
			\includegraphics[scale=0.3500]{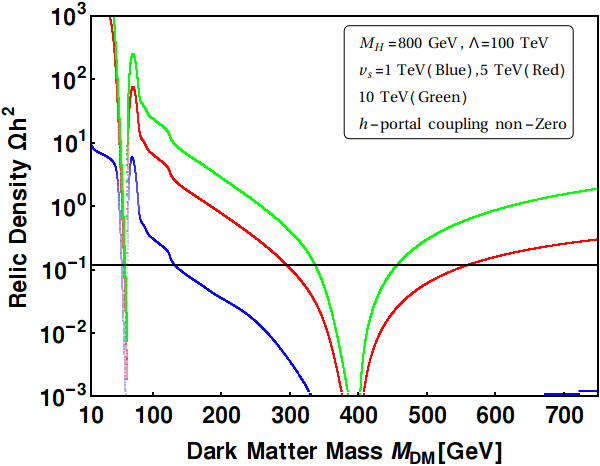}}	
\caption{ \it We varied the DM mass for three different singlet scalar VEVs. The Higgs $h$ portal coupling is non-zero here. The blue line, corresponding to the singlet scalar VEV $v_s=1$ TeV, is excluded by direct detection data from Xenon-1T~\cite{Aprile:2018dbl} and LUX-ZEPLIN (LZ)~\cite{LZ:2022ufs}. 
The red and green lines represent $v_s=5$ and $10$ TeV, respectively, and these are allowed from all the constraints, as discussed in the text.
The black line denotes the $3\sigma$ relic density band $\Omega h^2$ $=0.1198\pm0.0012$. }
		\label{fig:DMPlotkinmix1}
	\end{center}
\end{figure}
\begin{table*}[h!]
\begin{center}\scalebox{0.8}{
	\begin{tabular}{|p{2.0cm}|p{1.25cm}|p{1.2cm}|p{1.2cm}|p{1.2cm}|p{1.4cm}|p{2.5cm}|p{4.6cm}|}
		\hline
		\hline
		Sl. No. &Mixing $\cos\alpha$& $M_{\rm DM}$ (GeV) & $M_{H_F}$ (GeV) & $v_s$ (TeV)& $\Omega_{\rm DM}h^2$&DD cross-sec (${\rm cm^3/s}$)&~~~~Contributions \\
		\hline
	    \hline
		&&&&&&&$\sigma(A_F \,A_F\rightarrow b \overline{b})~~76\%$\\
		BP-1b	&~0.995&~54.75&~800&1.0&0.1196&$2.45 \times 10^{-45}$& $\sigma(A_F \,A_F\rightarrow q \overline{q})\quad 6 \%$\\
		 &&&&&&&$\sigma(A_F \,A_F\rightarrow l \overline{l})~~4\%$\\
		&&&&&&&$\sigma(A_F \,A_F\rightarrow W W^* )~~13\%$\\
		&&&&&&&$\sigma(A_F \,H^\pm\rightarrow ZZ^*)~~1\%$\\
		\hline
	    \hline
		&&&&&&&$\sigma(A_F \,A_F\rightarrow b \overline{b})~~76\%$\\
		BP-2b	&~0.995&~58.25&~800&5.0&0.1170&$5.65 \times 10^{-48}$& $\sigma(A_F \,A_F\rightarrow q \overline{q})\quad 6 \%$\\
		 &&&&&&&$\sigma(A_F \,A_F\rightarrow l \overline{l})~~4\%$\\
		&&&&&&&$\sigma(A_F \,A_F\rightarrow W W^* )~~12\%$\\
		&&&&&&&$\sigma(A_F \,H^\pm\rightarrow ZZ^*)~~2\%$\\
		\hline
	    \hline
		&&&&&&&$\sigma(A_F \,A_F\rightarrow b \overline{b})~~76\%$\\
		BP-3b	&~0.995&~62.757&~800&10.0&0.1145&$4.63 \times 10^{-49}$& $\sigma(A_F \,A_F\rightarrow q \overline{q})\quad 6 \%$\\
		 &&&&&&&$\sigma(A_F \,A_F\rightarrow l \overline{l})~~4\%$\\
		&&&&&&&$\sigma(A_F \,A_F\rightarrow W W^* )~~12\%$\\
		&&&&&&&$\sigma(A_F \,H^\pm\rightarrow ZZ^*)~~2\%$\\
		\hline
		\hline
	\end{tabular}}
\end{center}
	\caption{\it DM annihilation contributions in the low mass region for a mixing angle $\cos\alpha=0.995$.}
	\label{tabDM:anni1a}
\end{table*}

When considering a small Higgs-portal coupling, denoted as $\lambda_3 \neq 0$, it becomes possible to get the precise relic density near the vicinity of the Higgs ($h$) mass resonance region.  This region is also allowed from all other constraints.  We illustrate this scenario in Figure~\ref{fig:DMPlotkinmix1}(a) and \ref{fig:DMPlotkinmix1}(b) for three distinct values of the singlet scalar VEV: $v_S=1$ TeV, $v_S=5$ TeV and $v_S=10$ TeV, respectively.  In Figure~\ref{fig:DMPlotkinmix1}(a), we consider a heavy CP-even flavon $H_F$ with a mass of 500 GeV while in Figure~\ref{fig:DMPlotkinmix1}(b) the mass of the CP-even flavon $H_F$ is 800 GeV. 
The blue line, corresponding to $v_s=1$ TeV, is excluded by Xenon-1T~\cite{Aprile:2018dbl} and LUX-ZEPLIN (LZ) ~\cite{LZ:2022ufs} data, due to the presence of a non-zero coupling $g_{A_F A_F H_F}$, which is proportional to $\lambda_2 v_s$ and non-zero fermionic coupling $g_{H_F \bar{f}f}\approx 2\cos\alpha \,Z_{ij}\, \frac{ v_{\rm SM}}{\sqrt{2} \,v_s}$.  In contrast, the other two lines, namely the red one (representing $v_s=5$ TeV) and the green one (representing $v_s=10$ TeV), remain consistent with all the constraints imposed by various considerations. It is to be noted that the parameter $Z_{ij}=\delta_{ij} \hat{y}^f$, i.e., an off-diagonal mixing, is taken to be zero (see Eq.~(\ref{eq:yuk1})).  We show 6 BPs (3 for each $M_{H_F}$) for the low DM mass region, including corresponding contributions to the DM relic density and direct detection cross-sections in Table~\ref{tabDM:anni1}--\ref{tabDM:anni1a}. However,  BP-1 with $v_s=1$ TeV is ruled out due to the present direct detection data. Then BP-2 and BP-3 with $v_s=5$ and $10$ TeV, respectively,  give an allowed relic density.  One can also see that the $A_F \, A_F \rightarrow b \overline{b}$ contribution for the low DM mass region is around $\sim 80\%$.
We also present a few BPs in Table~\ref{tabDM:anni2a}--\ref{tabDM:anni2b} for heavier DM masses where the contributions dominantly come from the $A_F \, A_F\rightarrow t \overline{t}, WW, hh, ZZ$ channels (see Figure~\ref{fig:DMannidiag}). The $A_F \, A_F \rightarrow t \overline{t}$ contribution dominantly comes from the point vertex and via the $H_F$-mediated s-channel. The coupling of the vertex $A_F \,A_F t \overline{t}$ is proportional to $v_{\rm SM}/\sqrt{2} v_s^2$ and 
$H_F A_F A_F$ is proportional to $v_{\rm SM}/\sqrt{2} v_s$, respectively (see Eq.~(\ref{eq:fermiYukfinal})). Hence, the $A_F \,A_F \rightarrow t \overline{t}$ contribution decreases with larger $v_s$. The contributions  $A_F \, A_F  \rightarrow WW, hh, ZZ$ appear due to the mixing between the CP-even scalar fields in this model. These contributions become zero if we dial out this mixing. We will discuss these scenarios in the following.
\begin{table*}[h!]
\begin{center}\scalebox{0.8}{
	\begin{tabular}{|p{2.0cm}|p{1.25cm}|p{1.2cm}|p{1.2cm}|p{1.2cm}|p{1.4cm}|p{2.5cm}|p{4.6cm}|}
		\hline
		\hline
		Sl. No. &Mixing $\cos\alpha$& $M_{\rm DM}$ (GeV) & $M_{H_F}$ (GeV) & $v_s$ (TeV)& $\Omega_{\rm DM}h^2$&DD cross-sec (${\rm cm^3/s}$)&~~~~Contributions \\
		\hline
	    \hline
		&&&&&&&$\sigma(A_F \,A_F\rightarrow t \overline{t})~~58\%$\\
		BP-3a	&~0.995&~313.65&~500&5.0&0.1198&$2.86 \times 10^{-48}$& $\sigma(A_F \,A_F\rightarrow h h)\quad 12 \%$\\
		&&&&&&&$\sigma(A_F \,A_F\rightarrow W W^* )~~20\%$\\
		&&&&&&&$\sigma(A_F \,H^\pm\rightarrow ZZ^*)~~10\%$\\
		\hline
	    \hline
		&&&&&&&$\sigma(A_F \,A_F\rightarrow t \overline{t})~~48\%$\\
		BP-4a	&~0.995&~561.34&~800&5.0&0.1196&$1.86 \times 10^{-46}$& $\sigma(A_F \,A_F\rightarrow hh)\quad 14 \%$\\
		&&&&&&&$\sigma(A_F \,A_F\rightarrow W W^* )~~26\%$\\
		&&&&&&&$\sigma(A_F \,H^\pm\rightarrow ZZ^*)~~13\%$\\
		\hline
		\hline
	\end{tabular}}
\end{center}
	\caption{\it DM annihilation contributions in the high mass region for a mixing angle $\cos\alpha=0.995$.}
	\label{tabDM:anni2a}
\end{table*}
\begin{table*}[h!]
\begin{center}\scalebox{0.8}{
	\begin{tabular}{|p{2.0cm}|p{1.25cm}|p{1.2cm}|p{1.2cm}|p{1.2cm}|p{1.4cm}|p{2.5cm}|p{4.6cm}|}
		\hline
		\hline
		Sl. No. &Mixing $\cos\alpha$& $M_{\rm DM}$ (GeV) & $M_{H_F}$ (GeV) & $v_s$ (TeV)& $\Omega_{\rm DM}h^2$&DD cross-sec (${\rm cm^3/s}$)&~~~~Contributions \\
		\hline
	    \hline
		&&&&&&&$\sigma(A_F \,A_F\rightarrow t \overline{t})~~30\%$\\
		BP-3b	&~0.995&~271.255&~500&10.0&0.1197&$1.04 \times 10^{-48}$& $\sigma(A_F \,A_F\rightarrow h h)\quad 21 \%$\\
		&&&&&&&$\sigma(A_F \,A_F\rightarrow W W^* )~~33\%$\\
		&&&&&&&$\sigma(A_F \,H^\pm\rightarrow ZZ^*)~~16\%$\\
		\hline
	    \hline
		&&&&&&&$\sigma(A_F \,A_F\rightarrow t \overline{t})~~22\%$\\
		BP-4b	&~0.995&~456.91&~800&10.0&0.1195&$2.91 \times 10^{-46}$& $\sigma(A_F \,A_F\rightarrow hh)\quad 21 \%$\\
		&&&&&&&$\sigma(A_F \,A_F\rightarrow W W^* )~~38\%$\\
		&&&&&&&$\sigma(A_F \,H^\pm\rightarrow ZZ^*)~~19\%$\\
		\hline
		\hline
	\end{tabular}}
\end{center}
	\caption{\it DM annihilation contributions in the high mass region for a mixing angle $\cos\alpha=0.995$.}
	\label{tabDM:anni2b}
\end{table*}

\begin{figure}[h!]
	\begin{center}
		\subfigure[]{
			\includegraphics[scale=0.3500]{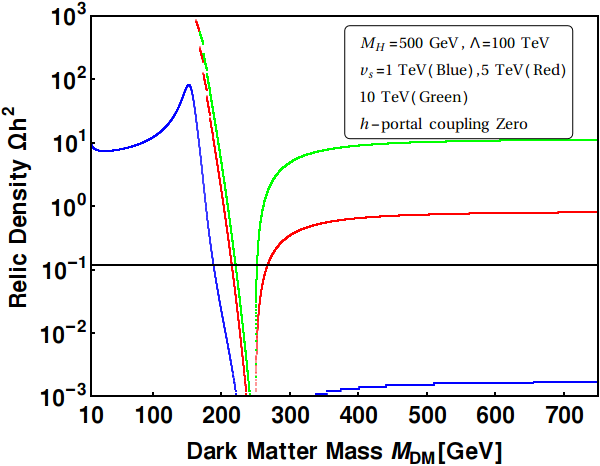}}	
		\subfigure[]{
			\includegraphics[scale=0.3500]{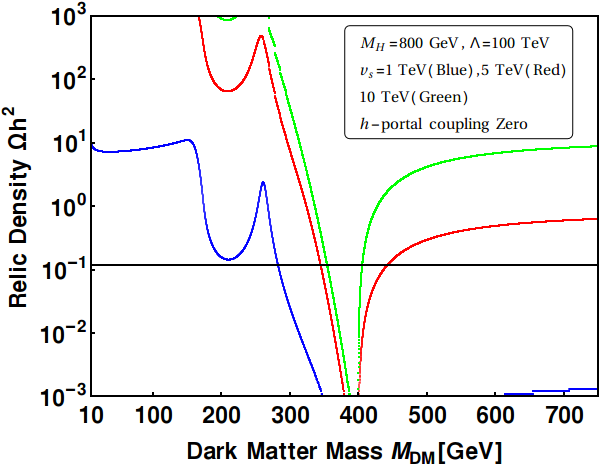}}	
		\caption{ \it We varied the DM mass for three different singlet scalar VEVs. The Higgs $h$ portal coupling is zero here, i.e., the mixing angle $\alpha=0$. The green and red lines represent $v_s=5$ and $10$ TeV, respectively, and these are allowed from all the constraints, as discussed in the text. The blue line, corresponding to $v_s=1$ TeV, is excluded by direct detection data from Xenon-1T~\cite{Aprile:2018dbl} and LUX-ZEPLIN (LZ)~\cite{LZ:2022ufs}.}
		\label{fig:DMPlotkinmix2}
	\end{center}
\end{figure}

Suppose we assume an entirely vanishing  Higgs portal coupling $g_{A_F A_F h}=0$, i.e., $\lambda_3=0$, which implies $\cos\alpha=0$. In that case, it becomes possible to obtain a relic density consistent with observations only for DM masses greater than $100$ GeV. 
In this case, the Yukawa sector can give an exact relic density. The corresponding contribution mainly comes via the $A_F \, A_F \rightarrow t \overline{t}$ channel (via a point vertex and the $H_F$-mediated $s$-channel).
We illustrate this in Figure~\ref{fig:DMPlotkinmix2} for three different VEVs, $v_s=1,5$ and $10$ TeV, respectively.  Figure~\ref{fig:DMPlotkinmix2}(a) is plotted with $M_{H_F}=500$ GeV, while Figure~\ref{fig:DMPlotkinmix2}(b) with $M_{H_F}=800$ GeV. All these lines (blue, red, and green) give exact and allowed relic density. We show a few BPs and corresponding contributions in Table~\ref{tabDM:anni3a}--\ref{tabDM:anni3b}.
\begin{table*}[h!]
\begin{center}\scalebox{0.8}{
	\begin{tabular}{|p{2.0cm}|p{1.25cm}|p{1.2cm}|p{1.2cm}|p{1.2cm}|p{1.4cm}|p{2.5cm}|p{4.6cm}|}
		\hline
		\hline
		Sl. No. &Mixing $\cos\alpha$& $M_{\rm DM}$ (GeV) & $M_{H_F}$ (GeV) & $v_s$ (TeV)& $\Omega_{\rm DM}h^2$&DD cross-sec (${\rm cm^3/s}$)&~~~~Contributions \\
		\hline
	    \hline
		BP-5a	&~1.0&~188.9&~500&1.0&0.1150&$1.06 \times 10^{-46}$&$ \sigma(A_F \,A_F\rightarrow t \overline{t})~~100 \%$\\
		\hline
	    \hline
		BP-6a	&~1.0&~215.30&~500&5.0&0.1192&$1.49 \times 10^{-49}$&$ \sigma(A_F \,A_F\rightarrow t \overline{t})~~100 \%$\\
		\hline
	    \hline
		BP-7a	&~1.0&~252.16&~500&10.0&0.1198&$7.8 \times 10^{-51}$&$ \sigma(A_F \,A_F\rightarrow t \overline{t})~~100 \%$\\
		\hline
		\hline
	\end{tabular}}
\end{center}
	\caption{\it DM annihilation contributions in the high mass region for $M_{H_F}=500$ GeV and a mixing angle $\alpha=0$.}
	\label{tabDM:anni3a}
\end{table*}
\begin{table*}[h!]
\begin{center}\scalebox{0.8}{
	\begin{tabular}{|p{2.0cm}|p{1.25cm}|p{1.2cm}|p{1.2cm}|p{1.2cm}|p{1.4cm}|p{2.5cm}|p{4.6cm}|}
		\hline
		\hline
		Sl. No. &Mixing $\cos\alpha$& $M_{\rm DM}$ (GeV) & $M_{H_F}$ (GeV) & $v_s$ (TeV)& $\Omega_{\rm DM}h^2$&DD cross-sec (${\rm cm^3/s}$)&~~~~Contributions \\
		\hline
	    \hline
		BP-5b	&~1.0&~282.70&~800&1.0&0.1192&$8.1 \times 10^{-47}$&$ \sigma(A_F \,A_F\rightarrow t \overline{t})~~100 \%$\\
		\hline
	    \hline
		BP-6b	&~1.0&~344.88&~800&5.0&0.1198&$7.54 \times 10^{-50}$&$ \sigma(A_F \,A_F\rightarrow t \overline{t})~~100 \%$\\
		\hline
	    \hline
		BP-7b	&~1.0&~354.48&~800&10.0&0.1198&$4.35 \times 10^{-51}$&$ \sigma(A_F \,A_F\rightarrow t \overline{t})~~100 \%$\\
		\hline
		\hline
	\end{tabular}}
\end{center}
	\caption{\it DM annihilation contributions in the high mass region  for $M_{H_F}=800$ GeV and a mixing angle $\alpha=0$.}
	\label{tabDM:anni3b}
\end{table*}

\begin{figure}[h!]
	\begin{center}
		\subfigure[]{
			\includegraphics[scale=0.400]{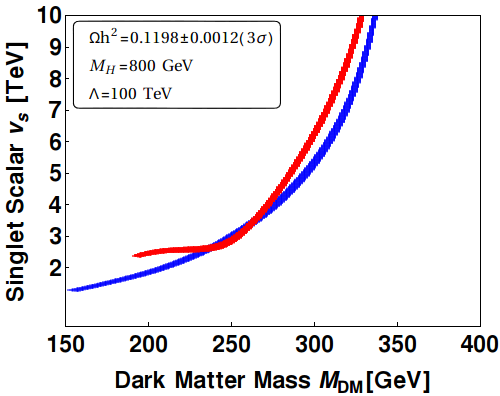}}	
		\caption{ \it We varied the DM mass and singlet scalar VEV.   The blue region shows the allowed DM relic density within the $3\sigma$ range. The region is plotted for  $Z_{L_i\ell_j}=\delta_{ij} \hat{y}^\ell$,~$Z_{Q_id_j}=\delta_{ij} \hat{y}^d$, and $Z_{Q_iu_j}=\delta_{ij} \hat{y}^u$. It also allowed from direct detection cross-section. We keep fixed $M_{H_F}=800$ GeV and mixing angle $\alpha=0$. The primary contributions to the DM relic density arise from the annihilation channels ($A_F A_F \to f\bar{f}$) through the Yukawa couplings as $\alpha=0$. }
		\label{fig:DMPlotkinmix3}
	\end{center}
\end{figure}

\begin{figure}[h!]
	\begin{center}
		\subfigure[]{
			\includegraphics[scale=0.400]{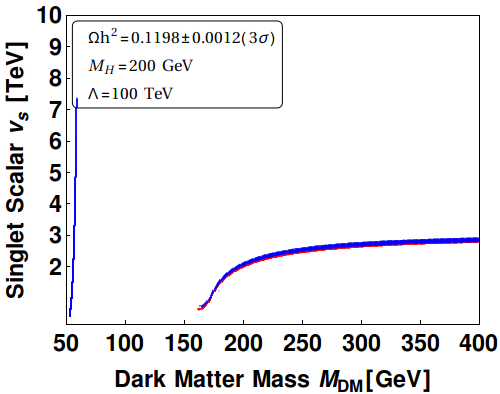}}	
		\caption{ \it We varied DM mass and singlet scalar VEV. We keep fixed $M_{H_F}=200$ GeV. Here $\cos\alpha=1$ for the red line, and $\cos\alpha=0.995$ for the blue line. 
  The red region shows the allowed DM relic density within the $3\sigma$ range. The red region is plotted for  $Z_{L_i\ell_j}=\delta_{ij} \hat{y}^\ell$,~$Z_{Q_id_j}=\delta_{ij} \hat{y}^d$, and $Z_{Q_iu_j}=\delta_{ij} \hat{y}^u$. }
		\label{fig:DMPlotkinmix4}
	\end{center}
\end{figure}

It is evident from these figures that the DM relic density is directly influenced by the value of the flavon breaking scale, i.e., singlet scalar VEV $v_s$ with or without the mixing of the CP-even scalars.
In the same context, we have depicted the regions in the DM mass vs. flavon VEV $v_s$ plane in Figure~\ref{fig:DMPlotkinmix3}, where the allowed DM relic density falls within the $3\sigma$ range.
We keep fixed CP-even flavon masses at $M_{H_F}=800$ GeV, the mixing parameter $\alpha=0$ allowed from the present LHC signal strength data.
The line in the figure is plotted under the assumption that the parameters $Z_{ij}$ are set as follows: $Z_{L_i\ell_j}=\delta_{ij} \hat{y}^\ell,~Z_{Q_id_j}=\delta_{ij} \hat{y}^d$, and $Z_{Q_iu_j}=\delta_{ij} \hat{y}^u$.
A lighter DM mass is allowed due to smaller coupling strength $A_F A_F h$ ($A_F A_F f\bar{f}$) is proportional to $Z_{ij}$ and inversely proportional to $v_s (v_s^2)$, see Eq.~(\ref{eq:fermiYukfinal}). It is noted that the DM mass also plays a role in the $g_{A_F A_F H_F}=(M_{{H_F}}^2+M_{{A_F}}^2)/v_s$ coupling. The large $v_s$ can reduce the effect of these $Z_{ij}$ and other coupling strengths. Hence, we get allowed DM mass for large $v_s$. In this case, we do not get allowed DM mass near the Higgs resonance region $M_{h}/2$ as  $g_{A_F A_F h}=0$. 
Allowing a small mixing angle, as characterized by our BP with $\cos\alpha=0.995$, facilitates achieving relic density in the vicinity of the Higgs resonance region for different values of $v_s$, contingent upon the specific $Z_{ij}$. The higher mass region remains largely unchanged as we have a slight variation of $\cos\alpha$. The same scenario we plotted in Figure~\ref{fig:DMPlotkinmix4} for another heavy Higgs mass $M_{H_F}=200$ GeV, keeping $Z_{ij}=1$ for all. The red line represents $\cos\alpha=1.0$, while the blue line represents $\cos\alpha=0.995$.

\begin{table*}[h!]
\begin{center}\scalebox{0.8}{
	\begin{tabular}{|p{2.0cm}|p{1.25cm}|p{1.2cm}|p{1.2cm}|p{1.2cm}|p{1.4cm}|p{2.5cm}|p{4.6cm}|}
		\hline
		\hline
		Sl. No. &Mixing $\cos\alpha$& $M_{\rm DM}$ (GeV) & $M_{H_F}$ (GeV) & $v_s$ (GeV)& $\Omega_{\rm DM}h^2$&DD cross-sec (${\rm cm^3/s}$)&~~~~Contributions \\
		\hline
	    \hline
	    &&&&&&&$ \sigma(A_F \,A_F\rightarrow b \overline{b})~~76 \%$\\
		BP-8a	&~1.0&~162.0&~200&625&0.1233&$2.40 \times 10^{-46}$&$ \sigma(A_F \,A_F\rightarrow t \overline{t})~~17 \%$\\
	    &&&&&&&$ \sigma(A_F \,A_F\rightarrow c \overline{c})~~6 \%$\\
	    &&&&&&&$ \sigma(A_F \,A_F\rightarrow \ell^+ \ell^-)~~4 \%$\\
		\hline
	    \hline
	    &&&&&&&$ \sigma(A_F \,A_F\rightarrow b \overline{b})~~36 \%$\\
		BP-9a	&~1.0&~168.0&~200&740&0.1186&$8.60 \times 10^{-47}$&$ \sigma(A_F \,A_F\rightarrow t \overline{t})~~59 \%$\\
	    &&&&&&&$ \sigma(A_F \,A_F\rightarrow c \overline{c})~~3 \%$\\
	    &&&&&&&$ \sigma(A_F \,A_F\rightarrow \ell^+ \ell^-)~~2 \%$\\
		\hline
	    \hline
	    &&&&&&&$ \sigma(A_F \,A_F\rightarrow b \overline{b})~~7 \%$\\
		BP-10a	&~1.0&~174.0&~200&1080&0.1135&$1.40 \times 10^{-47}$&$ \sigma(A_F \,A_F\rightarrow t \overline{t})~~92 \%$\\
		\hline
	    \hline
		BP-11a	&~1.0&~200.0&~200&2010&0.1189&$3.4 \times 10^{-76}$&$ \sigma(A_F \,A_F\rightarrow t \overline{t})~~100 \%$\\
		\hline
		\hline
	\end{tabular}}
\end{center}
	\caption{\it DM annihilation contributions high mass region for $M_{H_F}=200$ GeV and a mixing angle $\alpha=0$.}
	\label{tabDM:anni4a}
\end{table*}

In the depicted plot (red region) in Figure~\ref{fig:DMPlotkinmix4}, the primary contributions to the DM relic density arise from the annihilation channels ($A_F A_F \to f\bar{f}$) through the Yukawa couplings as the mixing angle $\alpha=0$. Depending on the DM mass, the contributions of different fermion final states are different. For the low DM mass, near $M_{\rm DM}=160$ GeV, $A_F A_F \to b\bar{b}$ the channel dominates. The contribution of $A_F A_F \to t\bar{t}$ increases with DM mass. We show such BPs in Table~\ref{tabDM:anni4a} for $M_{H_F}=200$ GeV, keeping $\alpha=0$ and $Z_{ij}=1$ for all.
It is worth noting that there are also partial contributions from other annihilation channels, e.g., $A_F A_F \to VV$, $V=W^\pm, Z$ for $\alpha>0$ as shown in Figure~\ref{fig:DMPlotkinmix4} (blue line).

\section{Collider Analysis }
\label{se:col_an}
Despite exhaustive efforts through collider and (in)direct detection experiments, searches for DM have yet to provide conclusive evidence of its existence. However, their dedication and expertise have significantly narrowed down many regions of phase space where the DM particle may lie hidden. The DM particle candidates generated in particle collisions remain undetectable by conventional collider experiments. Nonetheless, residual products from these collisions are observable. Invisible particles may be accompanied by one or more visible recoiling particles, resulting in a deficit of momentum in the transverse plane, referred to as $\slashed{E}_T$. It serves as a prominent signature of DM in collider experiments. In this model, we can get such kinds of events with large $\slashed{E}_T$ accompanied by (multiple) jets or leptons/photons in the final state. In the iFNSM framework, the monojet + $\slashed{E}_T~$ final state topology is found out to be an excellent probe of the DM sector; which we discuss in detail below. Here, we choose only one BP (among those discussed in the previous section) and perform a detailed collider analysis in the context of Run 3 of the LHC. We will perform a detailed cut-based analysis to identify a Signal Region (SR) that optimizes the signal detection while minimizing the SM backgrounds. Using this optimized SR, we will then analyze the variation in signal significance for different DM masses.

The largest production cross-section for the signal at the LHC arises from the gluon-gluon fusion process, contingent upon the mass of the DM candidate $A_F$. The region $M_{A_F} < M_h/2$ gives a large production cross-section through SM-like Higgs boson-mediated processes that could be excluded by the constraints on the invisible Higgs decay width~\cite{CMS:2022qva, ATLAS:2022tnm} (We ignore this region $M_{\rm DM}<M_h/2$). Hence, dominant production modes would be via heavy $H_F$ mediated processes due to large coupling between the heavy Higgs and the gluons via top-quark loops. 

\begin{figure}[htb!]
\begin{center}
\includegraphics[scale=0.6]{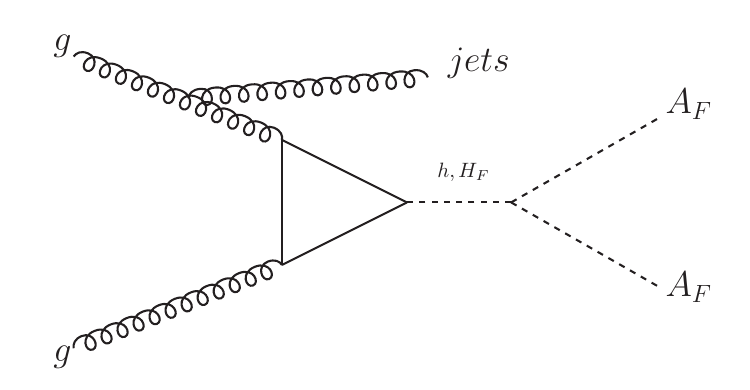}
\includegraphics[scale=0.6]{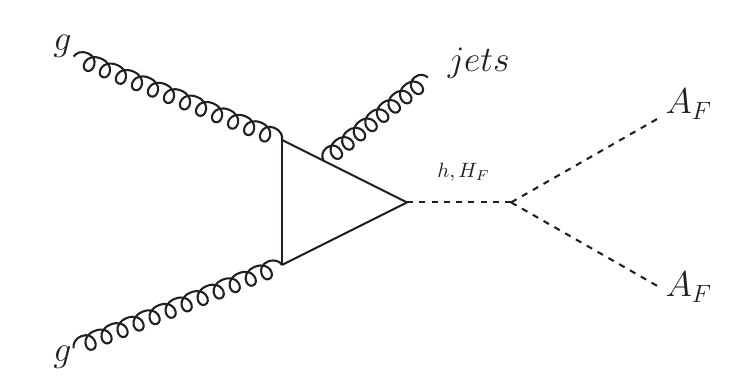} 
\end{center}
\caption{ \it Two representative Feynman diagrams that lead to mono-jet plus missing transverse energy in the final state. Each diagram captures the dominant production modes, with other subdominant modes considered as described in the accompanying text.}
\label{fig:crossDM}
\end{figure}

While simulating the signal events, we set $c_\alpha =0.995$ (i.e., a small mixing angle $\alpha$ between the CP-even part of the doublet and singlet scalar fields) and assume for the cut-off scale $\Lambda=100$ TeV, to satisfy theoretical as well as experimental bounds as discussed in the previous section. Note that, apart from the primary production mode, as illustrated in Figure~\ref{fig:crossDM}, numerous additional subdominant diagrams can generate an identical final state, such as $ q q \to A_F A_F + {\rm jets}$, where the jets originate from a $q$-leg. However, this channel is attenuated by singlet scalar VEVs $v_s^2$. Furthermore, VBF processes, mediated through Higgs bosons h/$H_F$, can also yield a similar final state, yet their production cross-section is suppressed by the small mixing angle $s_\alpha = \sqrt{1-c_\alpha^2} \sim 0.1$. Specifically, the vector boson fusion process mediated by $H_F$ is suppressed by $\mathcal{O}(s_\alpha^2)$ and $\frac{1}{v_s^2}$, while the $h$-mediated fusion is suppressed by $\mathcal{O}(s_\alpha^2)$, incorporating the Higgs boson mass squared term in the scalar propagator. Therefore, in this study, we focus on final state topology having an energetic jet with $p_T$ greater than 150 GeV along with large missing $\slashed{E}_T$ in the context of the 14 TeV run of LHC  with an integrated luminosity of 300 ${\rm fb}^{-1}$ (the aforementioned Run 3). 

We utilize {\tt FeynRules}~\cite{Alloul:2013bka} to construct the iFNSM model and generate the UFO files compatible with {\tt MadGraph-2.6.7}~\cite{Alwall:2014hca}. Subsequently, employing the particle spectrum associated with the BP, we simulate the events (at leading order in $\alpha_s$) for the loop-induced process $p p \to A_F A_F$ via spin-0 mediators $h$ and $H_F$ with one additional parton in the matrix element calculations (see Figure~\ref{fig:crossDM})\footnote{
The cross-section for other production modes of $h$ and $H_F$ with leptonic final states i.e., $p p \rightarrow H_F \, H_F, \,H_F \rightarrow \ell^+ \ell^-, \, H_F \rightarrow A_F A_F $ and $pp\rightarrow h H_F, \, h \rightarrow \ell^+ \ell^-, \, H_F \rightarrow A_F A_F $, where ($\ell = e,\mu, \tau$.) are found out to be insignificant.}. These events are then passed through \texttt{Pythia-8} \cite{Sjostrand:2014zea} for the parton showering and hadronization. The emulation of the detector response is accomplished using \texttt{Delphes-3.5.0} \cite{deFavereau:2013fsa} employing the default ATLAS configuration card. The jets are reconstructed using \texttt{Fastjet} \cite{Cacciari:2011ma} with the Tower information with jet radius parameter R = 0.6 and anti-$k_T$ jet algorithm \cite{Cacciari:2008gp} with minimum transverse momentum set at 20 GeV and pseudo-rapidity $|\eta| < 3$.

\begin{table}[!htb]
\begin{center}\scalebox{0.85}{
\begin{tabular}{|c|c|c|c|c|c|c|c|c|c|c|c|c|c|}
\hline
BP  & cross-section [pb] \\ [2mm]
\hline
\rule{0pt}{1ex}
$M_{H_F}=300$ GeV, \quad $M_{A_F}=100$  GeV &   0.456  \\  [2mm]
\hline
\end{tabular}}
\end{center}
\caption{ \it The production cross-section for the process $pp \to H_F\to A_F A_F  + {\rm jets}$. We fix $ n_{ij}=\delta_{ij}$ and  $M_{h}=125.5$ GeV, $\cos\alpha=0.995, v_s=1.5$  TeV and 
$\Lambda=100$ TeV. Note, the production cross-section is well within the current limit obtained from the LHC measurements \cite{ATLAS:2021kxv,CMS:2021far}.}
\label{tab:BPs}
\end{table}

Several SM processes can contribute as background to the mono-jet + $\slashed{E}_T$ signature. Among these, the $pp \to Z + {\rm jets}$ with $Z \to \sum_{\ell = e, \mu, \tau} \nu_\ell \bar{\nu_\ell}$ is the dominant one. Other processes like $pp \to W + {\rm jets}$ ($W \to \ell \nu_\ell$) , $pp \to VV + {\rm jets}$, $pp \to VVV \,+{\rm jets}$ (with $V=W^\pm,Z$), $pp \to t \bar{t} + {\rm jets}$ and QCD multijet processes also add to the backgrounds. 
For all of these processes, events are generated up to 2 partons at the matrix element level and subsequently MLM matching scheme \cite{Mangano:2006rw} is used to combine the jets from the matrix element calculations with the ones coming from parton shower. Additionally, a negligibly small contribution to the monojet + $\slashed{E}_T$ events can also arise from $pp \to Z^*, \gamma \to \ell^+ \ell^-  + {\rm jets}$ ($\ell=e,\mu,\tau$)
where the electrons/muons are faked as jets. The contribution of the top quark production process associated with additional gauge bosons is negligible and therefore not considered here. However, the contribution coming from most of these processes, except $pp \to Z + {\rm jets}$, eventually becomes negligible once the monojet selection cuts are imposed. The simulation of the SM background events follows the same procedure as that of the signal events. For both signal and background processes, we rely on Leading Order (LO) cross-sections, unless explicitly stated otherwise. We show the production cross-section for our reference BP (signal) and background processes in  Table~\ref{tab:BPs} and Table~\ref{tab:csBG}, respectively.

\begin{table}[htb!]
\begin{center}\scalebox{1.0}{
\begin{tabular}{|c|c|c|c|c|c|c|c|c|c|c|c|c|c|}
\hline
 SM backgrounds & cross-section [pb]  \\
\hline
\rule{0pt}{1ex}
$ pp \to Z + {\rm jets}  $ ($Z\to \nu_\ell \bar{\nu}_\ell$, $\ell=e,\mu,\tau$)&    $1.09\times 10^4$    \\
$pp \to W + {\rm jets}$ ($W \to \ell \nu_\ell$) &  93.9   \\
$ pp \to t \bar{t} + {\rm jets}$&    915.5    \\
$ pp \to W W + {\rm jets}  $ &    89.20   \\
$ pp \to W Z + {\rm jets}  $ &    40.10   \\
$ pp \to Z Z + {\rm jets} $ &    11.6   \\
$ pp \to VVV + {\rm jets} $  ($V=W,Z$)&    1.04\\   
\hline
\end{tabular}}
\end{center}
\caption{ \it The cross-sections for the most relevant SM background processes. }
\label{tab:csBG}
\end{table}

The events (both signal and background) are required to pass the following basic selection cuts on the reconstructed objects, namely the jets (also leptons and photons) and missing transverse energy, following \cite{ATLAS:2021kxv}: 

\begin{itemize} 
\item A leading jet with $p_T >$ 150 GeV and up to three additional jets with $p_T >$ 30 GeV. 

\item The missing transverse energy satisfy $\slashed{E}_T > $ 200 GeV. Additionally, the azimuthal angle separation between the missing energy and all the selected jets must satisfy $\Delta\phi({\rm jet}, \slashed{E}_T) > 0.5$ to reduce the QCD multijet background where large $\slashed{E}_T$ originates mainly from the mis-measurement of jet energy.

\item Leptons and photons having $p_T > 10$ GeV and $\mid \eta \mid < 2$ are vetoed.   
\end{itemize}

\begin{figure}[h!]
\begin{center}
\includegraphics[scale=0.27]{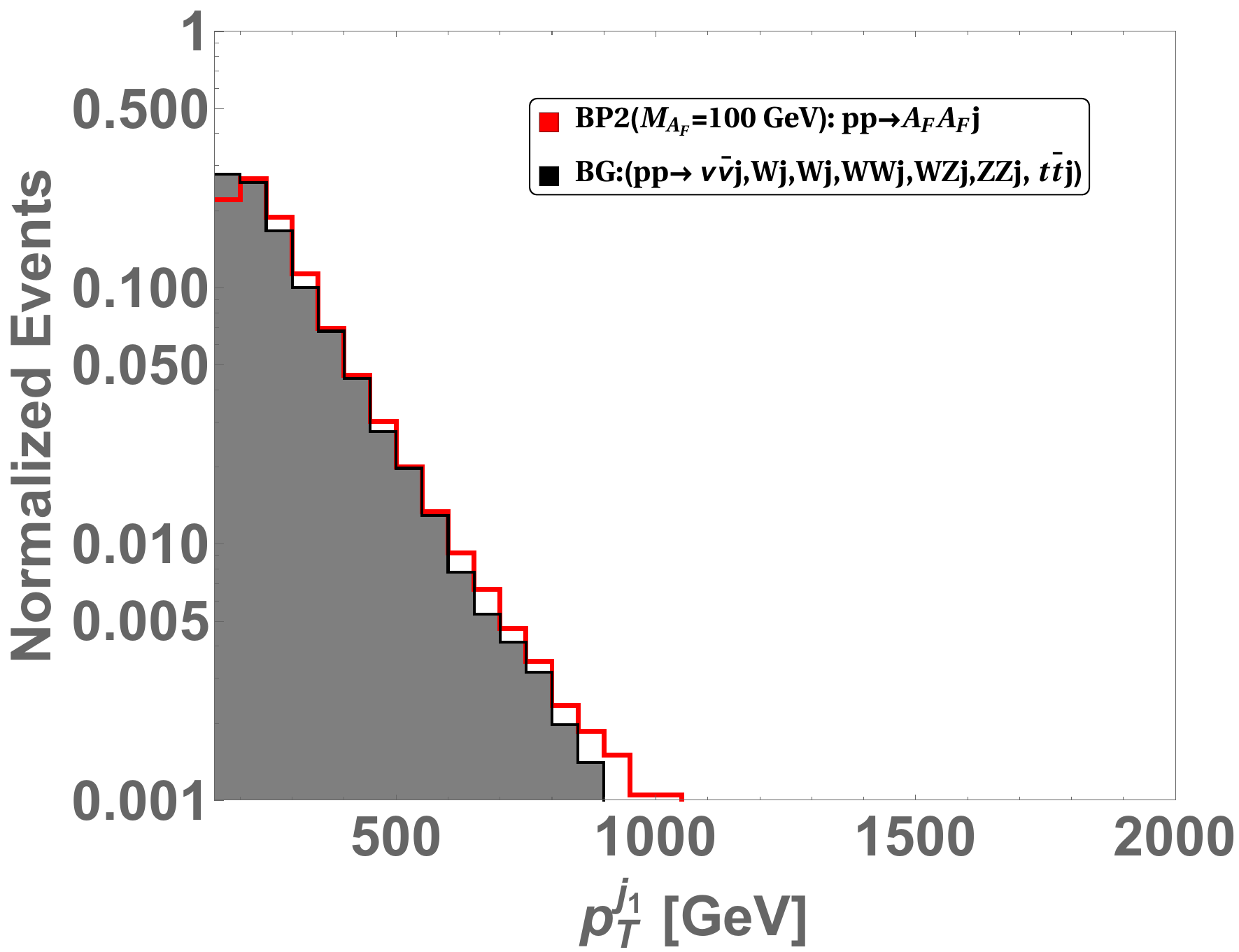} 
\includegraphics[scale=0.27]{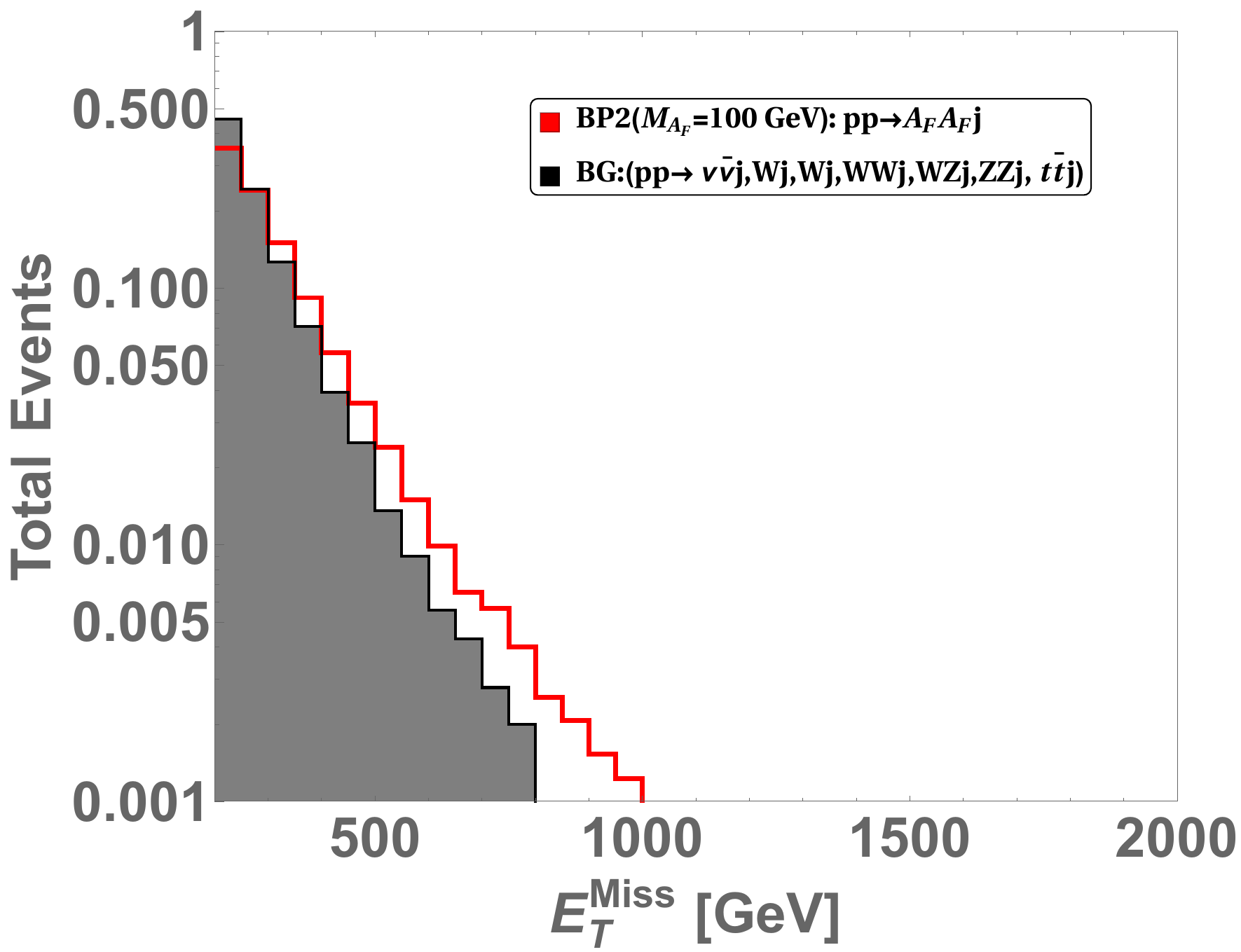}\\
\includegraphics[scale=0.27]{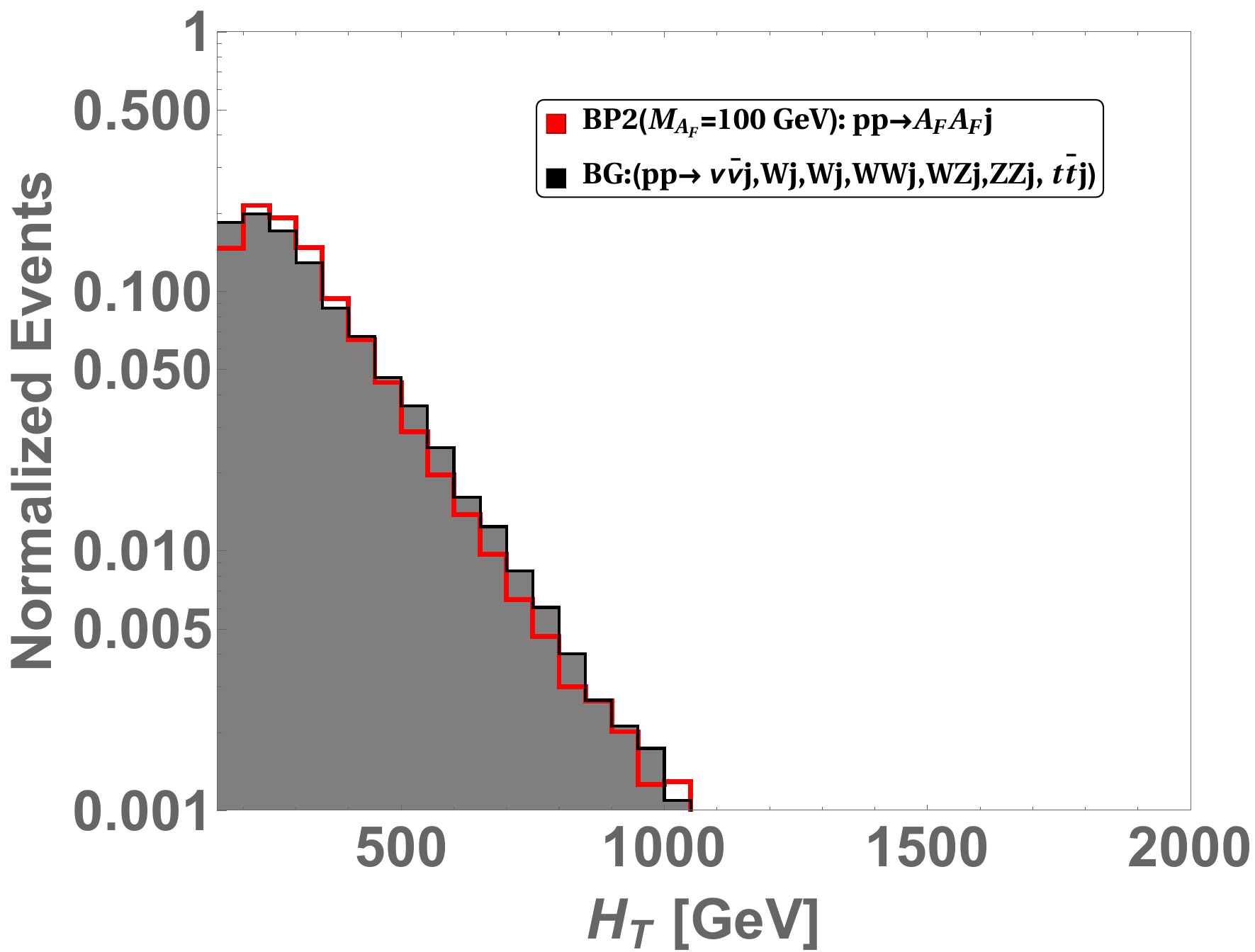}
\includegraphics[scale=0.27]{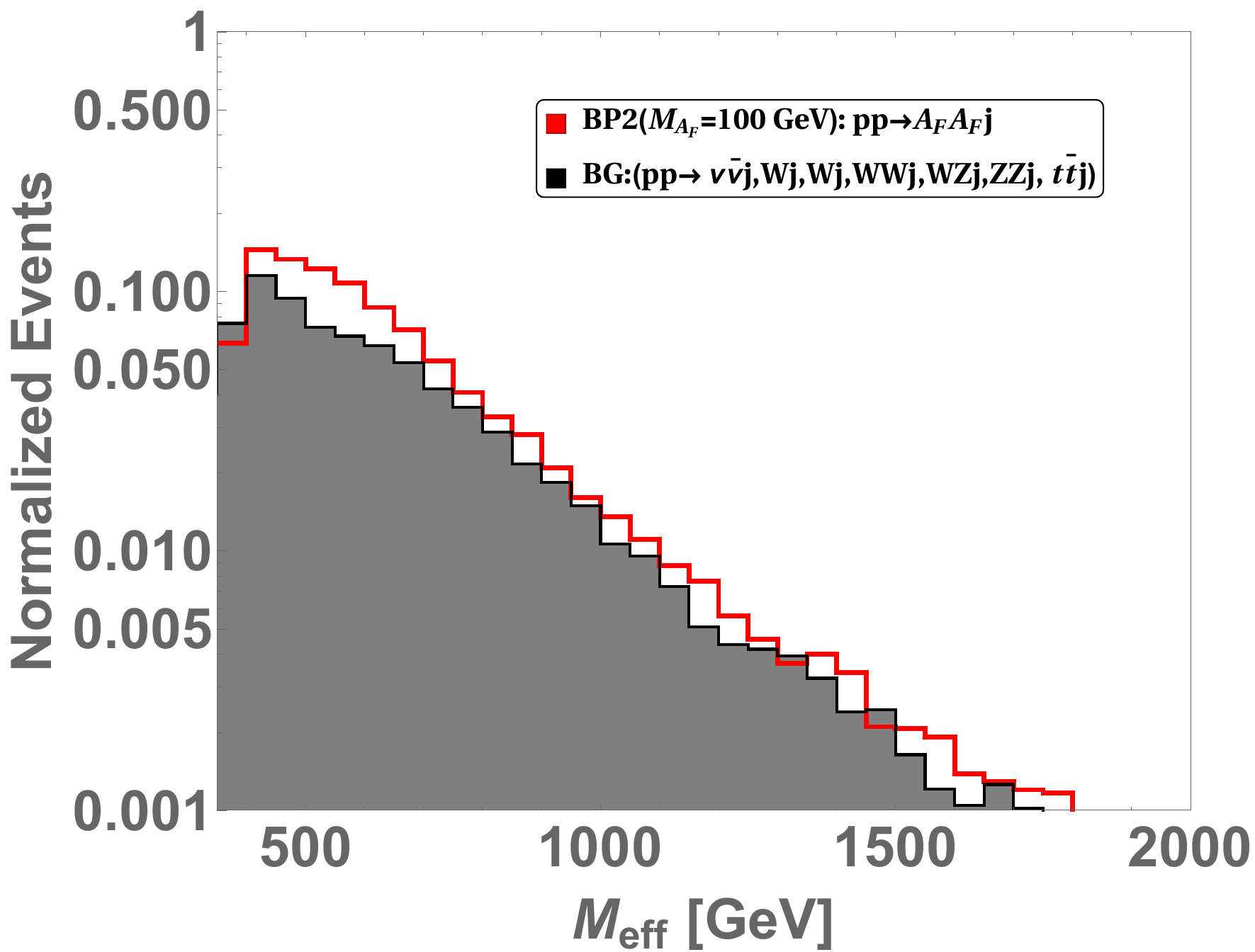}\\
\includegraphics[scale=0.27]{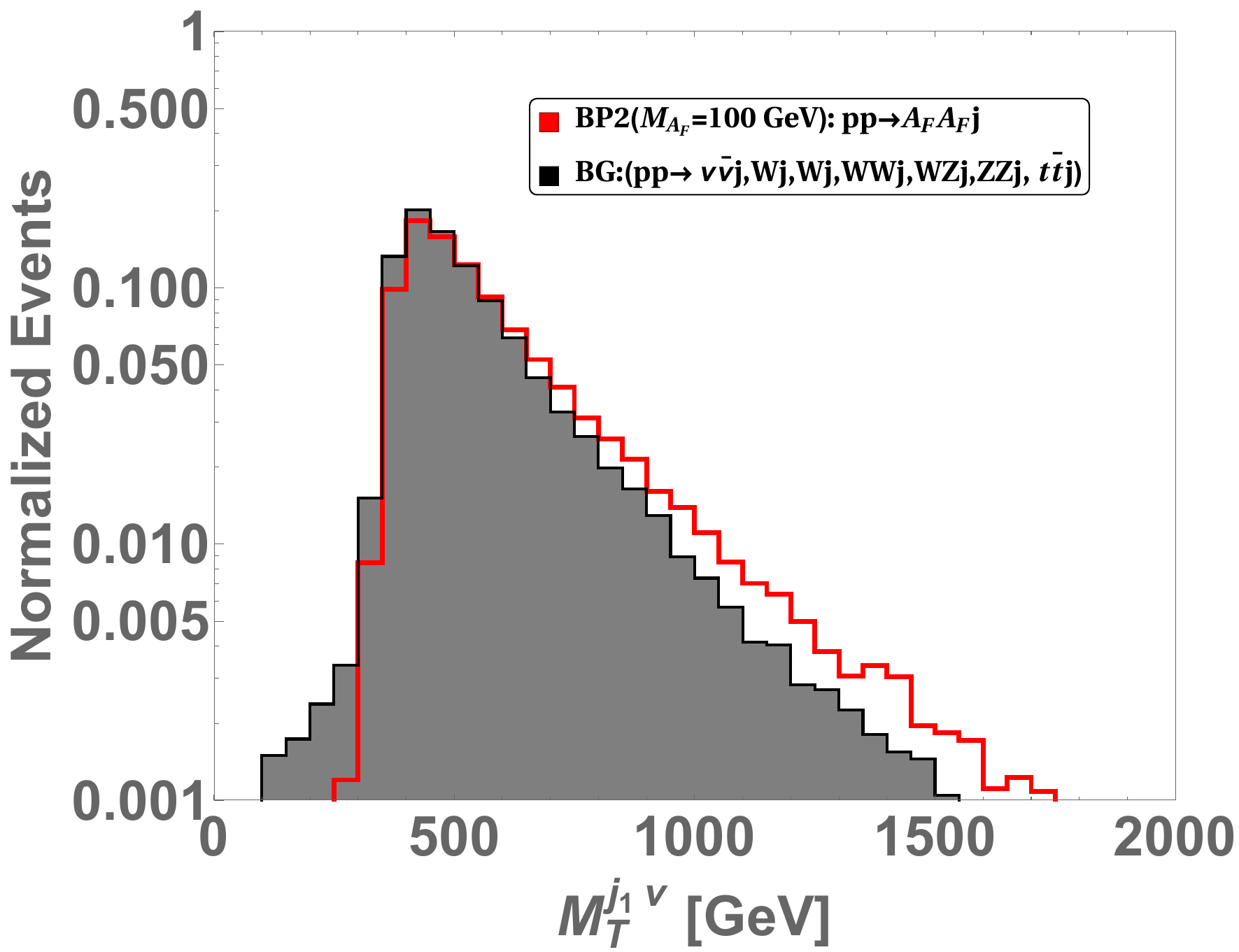}
\includegraphics[scale=0.27]{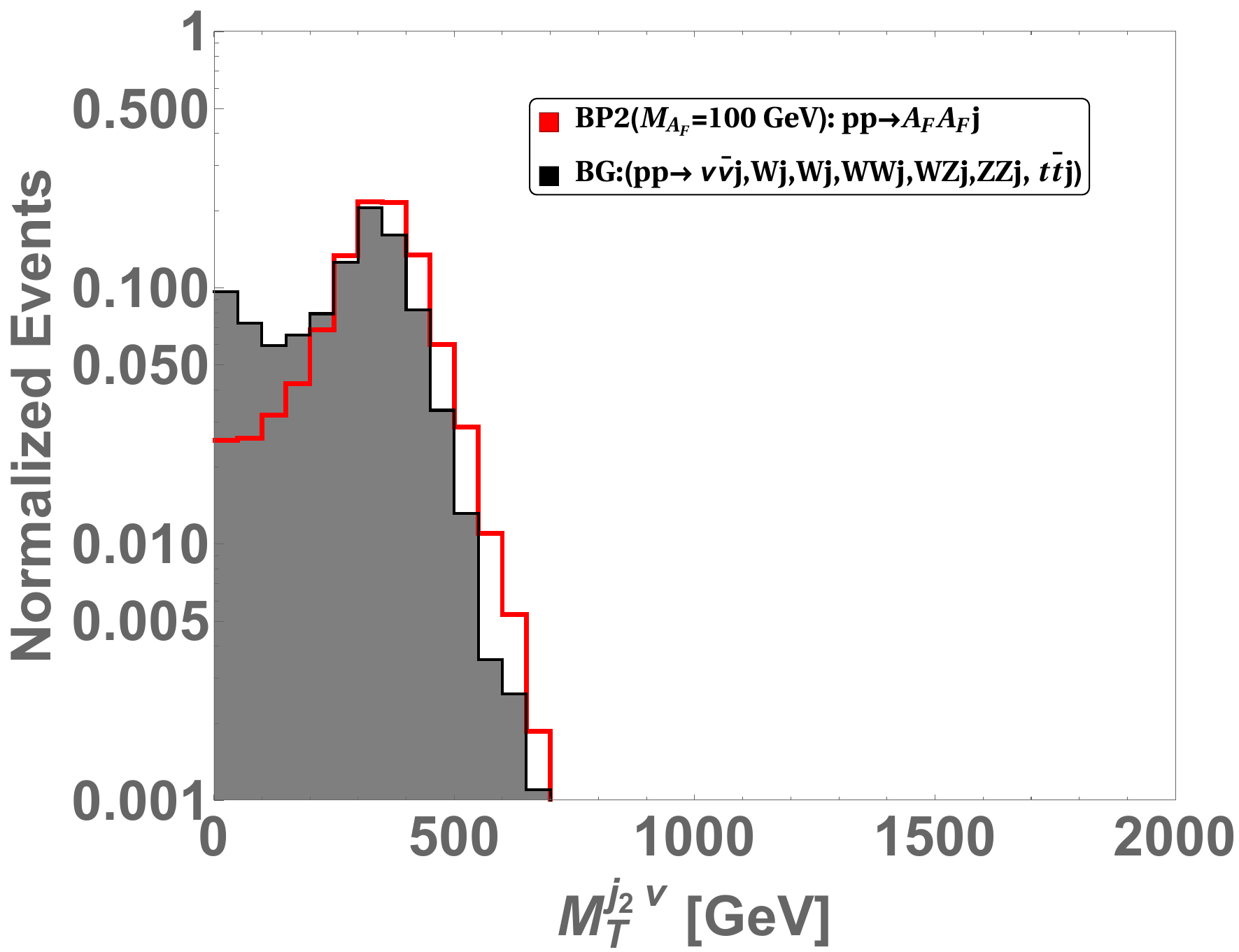}
\end{center}
\caption{ \it Normalized event distributions for signal and total background events after satisfying the basic acceptance cuts. The gray region takes into account the contribution of all the backgrounds weighted according to their cross-sections. }
\label{fig:Dist2}
\end{figure}

Once these basic requirements are satisfied, we then apply more stringent selection criteria using additional kinematic variables calculated using the reconstructed objects. This approach aims to bolster the signal-to-background ratio by identifying the phase space regions mostly populated by the signal events. First, we calculate the variable called $H_T$ which is defined as the scalar $p_T$ sum of all the visible objects, namely the jets. Next, we estimate the total energy associated with the process using the variable $m_{\rm eff}$ defined as $m_{\rm eff} = \sum_i p_T^i + \slashed{p}_T$, where $i$ runs over all the detector level objects. The peak around $500$ GeV observed in  $m_{\rm eff}$ distribution originates due to (scalar) addition of the contributions coming from $H_T$ and $\slashed{p}_T$, where for the later the events are selected with $\slashed{E}_T >$ 200 GeV. 
To minimize the contribution of leptons faking as jets, we also calculate the transverse mass using the leading two jets and the missing transverse energy, defined as $M_T^{j_i\nu}= \sqrt{p_T^{j_i} \slashed{E}_T (1- \cos \Delta \phi_{{j_i} \slashed{E}_T }) }$, here, $i=1, \, 2$, and $\Delta \phi_{{j_i} \slashed{E}_T }$ is the difference in azimuthal angle between the transverse momentum of the jets and missing transverse energy $\slashed{E}_T$. The normalized distribution of all these kinematic variables for both the signal and the combined background events are depicted in Figure~\ref{fig:Dist2}. After a thorough investigation of these distributions, we proceed with a comprehensive cut-based analysis to optimize the discovery potential of the signal events over the SM backgrounds. The selection cuts applied to those observables are shown in Table~\ref{table:sr}.

\begin{table}[h!]
\begin{center}\scalebox{1.0}{
\begin{tabular}{|c|c|c|c|c|c|c|c|c|c|c|c|c|c|}
\hline
  &\multicolumn{2}{c|}{ Kinematic variables and cuts   }\\
\cline{2-3}
& ~~~~Observable~~~~& ~~~~~~~~~Value~(GeV)~~~~~~~~\\
\cline{1-3}
\rule{0pt}{1ex}
\rule{0pt}{1ex}
  & $p_T^{j_1}$ & $>$ 150 ~~~~~~~~\\
\rule{0pt}{1ex}
SR~   & $\slashed{E}_T$ &  $>$ 285 ~~~~~~~~ \\
\rule{0pt}{1ex}
 & $M_T^{j_1\nu}$ &  $>$ 240  ~~~~~~~~\\
\hline
\end{tabular}}
\end{center}
\caption{ \it The optimized SR, which gave maximum significance mono-jet plus missing transverse energy analysis for our BP $M_{H_F}=300$ GeV, $ M_{A_F} (\equiv M_{\rm DM})=100$ GeV with $v_{v}=1.5$ TeV and $\Lambda=100$ TeV.}
\label{table:sr}
\vspace*{1cm}
\end{table}

\begin{figure}[htb!]
\begin{center}
\includegraphics[scale=0.27]{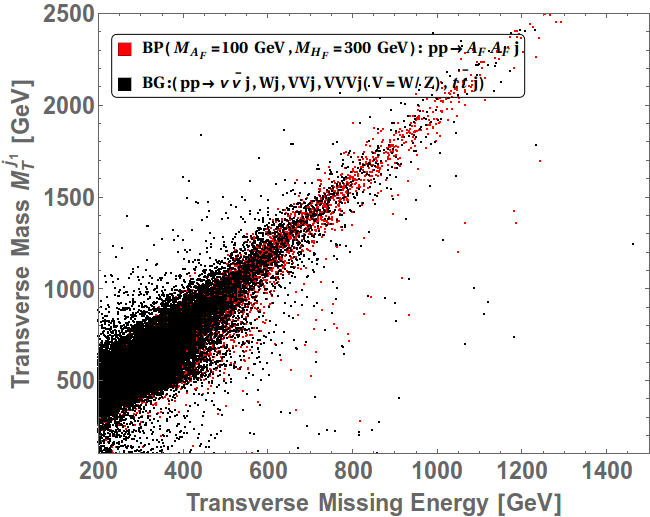} 
\includegraphics[scale=0.27]{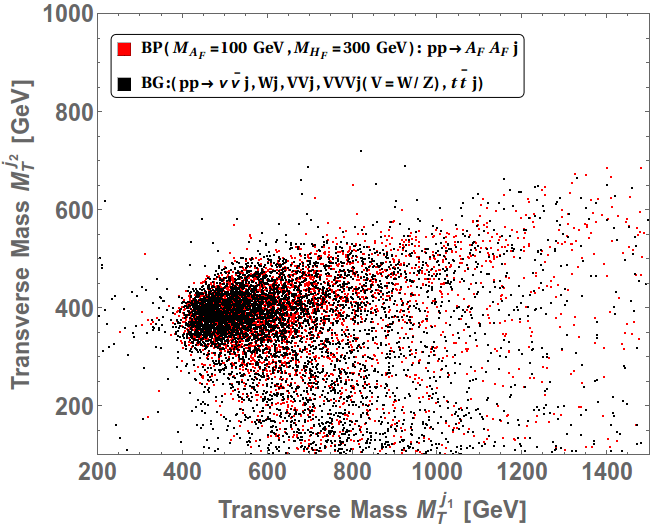}\\
\includegraphics[scale=0.27]{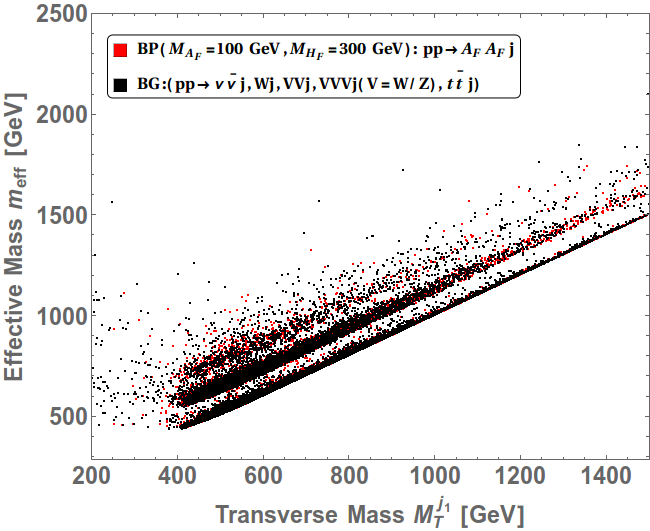}
\includegraphics[scale=0.27]{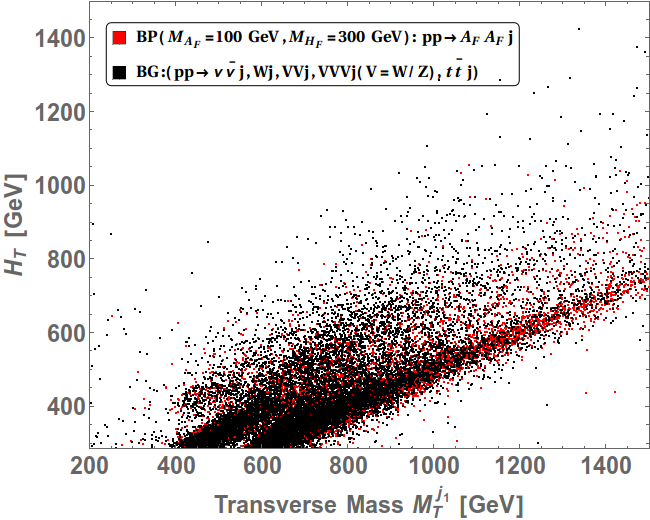}
\end{center}
\caption{ \it The correlation plots depict distinct scenarios. The first $p_T^{j_1}~{vs}~\slashed{E}_T$ plot (top-left) is generated after acceptance cuts. Conversely, the second (top-right) $M_T^{j_1}~{vs}~M_T^{j_2}$, third (left-bottom) $M_T^{j_1}~{vs}~m_{\rm eff}$ and fourth (right-bottom) $M_T^{j_1}~{vs}~H_T$ plots are produced subsequent to further imposing additional cut $\slashed{E}_T> 285$ GeV.}
\label{fig:correlation}
\end{figure}

The signal yields for the BP, along with the
corresponding background ones,  obtained after the 
application of the acceptance and selection cuts defining the SR as shown in Table~\ref{table:signalsignificance} for $\sqrt s=14$ TeV at $\mathcal{L}=$  $300 ~ {\rm fb^{-1}}$. The transverse momentum of the jets $p_T^{j_1}> 150$ GeV and missing transverse energy $\slashed{E}_T> 285$ GeV significantly suppresses background contributions as we have DM mass $100$ GeV. 
Incorporating additional cuts in the presence of $p_T^{j_1}$ and $\slashed{E}_T$, such as a transverse mass cut for the leading jet, $M_T^{j_1\nu}> 240$ GeV, further reduces the backgrounds.
It is to be noted that the cuts $M_T^{j_2\nu}, H_T, m_{\rm eff}$ are almost insensitive here after the imposition of the previous cuts.
We present such correlation plots to elucidate the sensitivity of these cuts in Figure~\ref{fig:correlation}, where the signal points are denoted in red, while SM background points are shown in black. The top-left panel illustrates the $\slashed{E}_T-M_T^{j_1}$ correlation plot generated post-application of acceptance cuts. Subsequently, we impose a cut on $\slashed{E}_T>285$ GeV and visualize the signal and background points in the $M_T^{j_1}-M_T^{j_2}$ (top-right) and $M_T^{j_1}-M_T^{j_2}$ (bottom) planes. It is discernible from these correlation plots that $M_T^{j_2\nu}$ and $m_{\rm eff}$ remain largely insensitive, primarily due to the $\slashed{E}_T>285$ GeV cut. We also check that additional cuts on $M_T^{j_2\nu}$ and/or $m_{\rm eff}$ do not improve the signal significance, as the signal and background distributions are largely overlapping. 

\begin{table}[h!]
\begin{center}\scalebox{1.0}{
\begin{tabular}{|cc|c|c|c|c|c|c|c|c|c|c|c|c|}
\hline
 & \multicolumn{2}{c|}{ Kinematic variables and cuts   } & \multicolumn{2}{c|}{ Total events after each cut } \\
\cline{1-5}
 &~~~~Observable~~~~~& ~~~~~~~~~Value~(GeV)~~~~~~~~&~~~~~ Background~~~~~& ~~~~~ Signal~~~~~\\
\cline{1-5}
\rule{0pt}{1ex}
\rule{0pt}{1ex}
&   Acceptance Cuts~ & -  &  $1.613 \times 10^8$     & 4.394$ \times 10^4$  \\
\rule{0pt}{1ex}
&   $p_T^{j_{1}}$ & $>$ 150  &   $2.994 \times 10^7$     &  $ 1.915\times 10^4$  \\
\rule{0pt}{1ex}
&   $\slashed{E}_T$ &  $>$ 285 &  $1.224 \times 10^6$     & $ 5.161 \times 10^3$  \\
\rule{0pt}{1ex}
&   $M_T^{j_1\nu}$ &  $>$ 240 &  $1.218 \times 10^6$     & $ 5.096 \times 10^3$ \\
\hline
\end{tabular}}
\end{center}
\caption{ \it The cut flow table after each cut, shown in Table~\ref{table:sr} at $\sqrt{s}=14$ TeV with integrated luminosity $\mathcal{L}=300~{\rm fb^{-1}}$ for our BP $M_{H_F}=300$ GeV, $ M_{A_F} (\equiv M_{\rm DM})=100$ GeV with $v_{v}=1.5$ TeV and $\Lambda=100$ TeV.}
\label{table:srCUT}
\vspace*{1cm}
\end{table}

A highly effective method to measure the efficacy of these cuts is to calculate the (statistical) signal significance. This is defined as~\cite{SigForm,Gross:2018okg} ${\mathcal S} = \sqrt{2 \left((S+B) \log \left(\frac{S+B}{B}\right)-S\right)}$, where $S$ and $B$ represent the number of signal and background events remaining after the selection cuts for a given integrated luminosity.
\begin{table}[htb!]
\begin{center}
{
\begin{tabular}{|c|c|c|c|c|c|c|c|c|c|c|c|c|c|}
\hline
~Total number of  background events~ & ~Total number of signal events~ &  ~Signal significance~ \\ [2mm]
\hline
$1.218 \times 10^6$ & $5.096 \times 10^3$ &   4.67  \\
\hline
\end{tabular}}
\end{center}
\caption{ \it The signal significance ${\mathcal S}$ for BP ($M_{H_F}=300$ GeV, $ M_{A_F} (\equiv M_{\rm DM})=100$ GeV with $v_{v}=1.5$ TeV and $\Lambda=100$ TeV.) corresponding to the optimized SR are shown. In addition, the total background yield and the total signal yield are also given at $\sqrt{s}=14$ TeV with integrated luminosity $\mathcal{L}=300~{\rm fb^{-1}}$.}
\label{table:signalsignificance}
\end{table}

In Table~\ref{table:srCUT}, we display the number of signals $S$ and background $B$ events surviving each cut at $\mathcal{L}=300~{\rm fb^{-1}}$. Both $S$ and $B$ are calculated by considering their respective production cross-sections and cut efficiencies.
We find the signal significance to be around 5$\sigma$ at $\sqrt{s}=14$ TeV with integrated luminosity $\mathcal{L}=300~{\rm fb^{-1}}$, as shown in Table~\ref{table:signalsignificance}.
This demonstrates that the sensitivity of the proposed signature at the HL-LHC remains largely unaffected by unknown factors influencing the background estimation, regardless of their origin.
An extensive upgrade program will be implemented for the LHC and its experiments in several phases~\cite{Apollinari:2015wtw}.
The High-Luminosity LHC (HL-LHC) is expected to deliver between $3000$ and $4000$ fb$^{-1}$ of data,
which is ten times more than the combined integrated luminosity of LHC Runs $1-3$.
According to our analysis, the signal significance for our BP could be enhanced from $4.67\sigma$ to $(14.5-16.6)\sigma$. It could be the most striking signature of new DM physics beyond the standard model.
\begin{figure}[ht!]
\begin{center}
\includegraphics[scale=0.27]{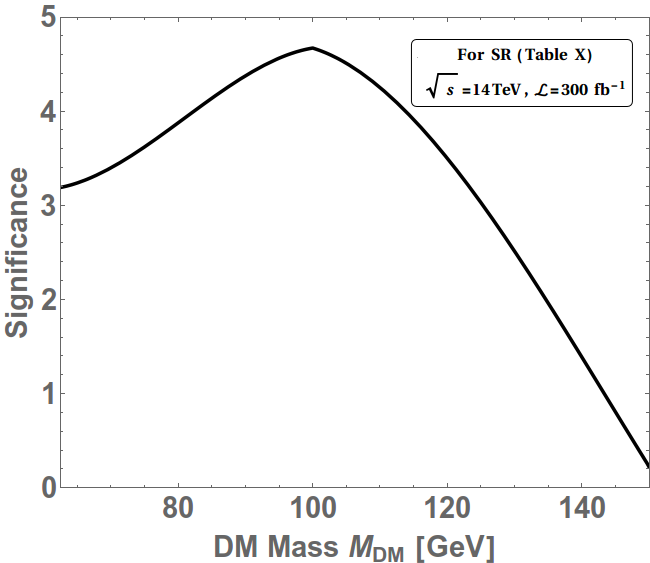} \includegraphics[scale=0.27]{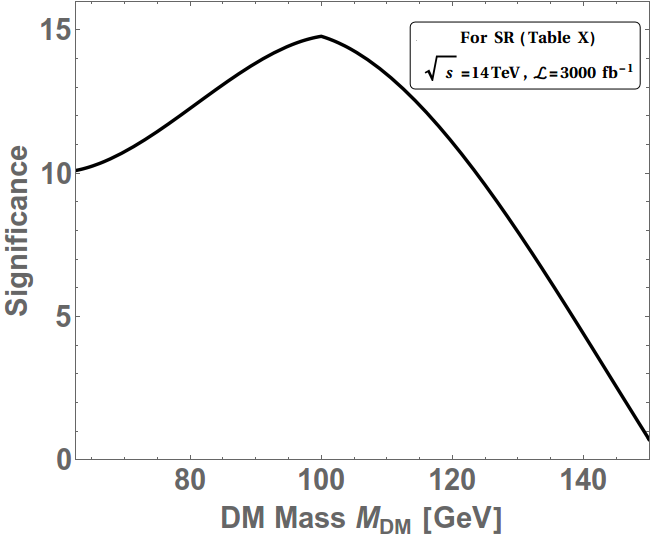} 
\end{center}
\caption{ \it The variation of signal significance $\mathcal{S}$ with DM mass. We use the same SR as in Table~\ref{table:sr}, i.e., $p_T^{j_1}> 150$, $\slashed{E}_T> 285$ and  $M_T^{j_1\nu}>240$. We generated this plot for $\sqrt{s}=14$ TeV, $\mathcal{L}=300~{\rm fb^{-1}}$ (left) and $\mathcal{L}=3000~{\rm fb^{-1}}$ (right).}
\label{fig:sigMass}
\end{figure}

We have generated additional data points to analyze how the significance varies with different DM masses, keeping the heavy flavon mass fixed at $M_{H_F} = 300$ GeV. The DM mass was varied from $M_h/2$ to $M_{H_F}/2$ with a step size of 20 GeV. The region below $M_h/2$ was avoided due to Higgs signal strength constraints. For DM masses above $M_{H_F}/2$, the cross-section is suppressed due to the non-resonant production of DM via $H_F$. We use the same SR as in Table~\ref{table:sr}, i.e., $p_T^{j_1}> 150$, $\slashed{E}_T> 285$ and  $M_T^{j_1\nu}>240$ to get the significance for various DM masses. We show such variation in Figure~\ref{fig:sigMass}. We generated two plots for $\sqrt{s}=14$ TeV with integrated luminosity $\mathcal{L}=300~{\rm fb^{-1}}$ and $\mathcal{L}=3000~{\rm fb^{-1}}$, respectively.

\section{Conclusions}
\label{sec:5}
In this study, we explored a scenario similar to the Froggatt-Nielson mechanism within the SM to explain the mass hierarchy, quark-mixing structure, and to investigate dark matter. Instead of using a continuous $U(1)_F$ symmetry, we employ discrete $\mathcal{Z}_2$ and $\mathcal{Z}_4$ symmetries to achieve these objectives. The iFNSM, a relatively unexplored framework in the realm of DM, offers a novel perspective and potential solutions to this particle physics enigma. Its unique features and underlying principles provide fertile ground for theoretical investigations into DM candidates and their interactions. Herein, we have scrutinized this DM model to unearth hidden theoretical connections and unveil viable phenomenological scenarios. 
A CP-odd (pNGB) particle emerging as an imaginary component of a flavon field in this scenario, which is neutral under the SM gauge group $SU(2)_L \times U(1)_Y$, can serve as a compelling DM candidate, in presence of the  unbroken $\mathcal{Z}_2$ symmetry, while the CP-even component of the same field emerges as the flavon itself, a configuration which defines our iFNSM.

Our analysis demonstrates that the mass of such a  DM candidate is intrinsically linked to the scale at which the $\mathcal{Z}_4$ (different from original $U(1)_F$ in FNSM~\cite{Froggatt:1978nt}) symmetry-breaking occurs via the relic density requirements. This linking is done via the coefficient of the $\frac{\left(S^2+ S^{*2}\right)}{\Lambda^2}$ Lagrangian term, that we  have introduced. Our analysis also reveals that the Yukawa couplings, which generate fermion masses within the FNSM, have a dual role. Not only do they contribute to the fermion mass hierarchy, but they also influence the properties of the DM candidate. This dual functionality suggests an intriguing connection between the DM sector and the generation of quark and lepton masses.

By considering the flavon VEV, quartic, and Yukawa couplings, we have identified a considerable allowed region of the iFNSM parameter space, representing a range of potential DM masses and coupling strengths consistent with all the theoretical and experimental constraints amenable to phenomenological investigation. Hence, this spurred us to conduct an in-depth analysis of this new BSM framework.

Within the latter, the  DM particle is subject to stringent constraints derived from stability, unitarity and perturbativity limits as well as those emerging from the SM Higgs boson coupling measurements and various searches targeting directly the DM particle itself. The detection of DM particles presents a unique challenge as they are impervious to direct observation, necessitating the presence of additional visible particles for detection. Hence, signatures such as mono-jet accompanied by significant missing transverse energy are frequently pursued. The mono-jet signature, with the hadronic activity originating from ISR, is particularly valuable due to its potential to alleviate dominant SM backgrounds through the imposition of a large transverse momentum requirement on the accompanying jet.

In our comprehensive study, we have evaluated the detectability of parameter space points that conform to relic density constraints and elude current (in)direct detection bounds on DM properties. In the context of the iFNSM, the primary mechanism for generating the DM particle ($A_F$) involves producing a heavy CP-even flavon $H_F$, followed by its subsequent decay into a pair of DM particles (i.e., $gg\to H_F\to A_F A_F$ accompanied by QCD activity). 
(We also checked that alternative detection channels, such as with leptonic final states, VBF production modes or associated production with a $Z$ or a Higgs bosons, prove less effective for the chosen BP.) Specifically, we have focused here on final states featuring at least one jet and large missing transverse energy in the context of the 14 TeV run of the LHC with an integrated luminosity of 300 ${\rm fb}^{-1}$ (Run 3). We have followed the procedure of a cut-based analysis to obtain a significance for our BP that can lead to its discovery, as it is found to be at $\sim 5\sigma$. 
In several phases of future analysis, an extensive upgrade program will be implemented for the LHC and its experiments. Chiefly, the HL-LHC is expected to deliver between $3000$ and $4000$ fb$^{-1}$ of data, which is ten times more than the combined integrated luminosity of LHC Runs 1-3. Herein, the signal significance for our BP could be enhanced from $\sim 5\sigma$ to between $\sim15\sigma$ and $17\sigma$. Thus, the CERN machine could potentially provide us with a most striking signature of DM.

In summary, our analysis of the iFNSM supports the possible detection of its DM candidate. Furthermore, it establishes a compelling link between its mass and the scale of flavon-symmetry breaking. This connection offers a unique perspective on the interplay between hidden and visible sectors of particle physics, potentially advancing our understanding of DM and the fermion mass hierarchy. Further experimental and observational efforts are therefore encouraged to explore and validate the predictions of this intriguing BSM scenario.

\subsection*{Acknowledgments}
NK sincerely thanks HRI-RECAPP (Prayagraj) for providing access to its cluster infrastructure and facilitating the computational analysis.
SM is supported in part through the NExT Institute and  STFC Consolidated Grant ST/L000296/1. 
The work of AC is funded by the Department of
Science and Technology, Government of India, under Grant No. IFA18-PH 224 (INSPIRE Faculty
Award).
\section{Appendix}
\subsection{Direct Detection Cross-section}
\label{eq:ddcross}
The spin-independent direct detection cross-section is given by:
\begin{eqnarray}
    \frac{d\sigma_{SI}}{d\Omega} = \frac{f_N^2 m_N^4}{16\pi (m_N+M_{\rm DM})^2}\, |\mathcal{M}_{\rm eff}|^2, 
\end{eqnarray}
where $d\Omega$ is the differential solid angle, the neucleon mass is $m_N=0.94$ GeV with the effective Higgs–nucleon coupling being  $f_N=0.3$~\cite{Alarcon:2012nr,Alarcon:2011zs,Cline:2013gha,Jiang:2023xdf}
Here, the effective square DM–nucleon amplitude due to these two scalar is:
\begin{eqnarray}
    |\mathcal{M}_{\rm eff}|^2  = \frac{1}{m_N^2 v_{\rm SM}^2 }\, \left[ \frac{g_{A_F A_F h} g_{N\overline{N} h} }{t- M_h^2}  + \frac{g_{A_F A_F H_F} g_{N\overline{N} H_F} }{t- M_{H_F}^2}   \right]^2\, (4m_N^2-t),
\end{eqnarray}
where, the coupling strength of the DM $A_F$ to the CP-even Higgses are $g_{A_F A_F h}$=$-i(\cos\alpha \, \lambda_{3} v_{\rm SM} $-$ 2 \sin\alpha\,  \lambda_{2} v_s )$=$i\,\sin\alpha\,(M_h^2+M_{{A_F}}^2)/v_s$ and $g_{A_F A_F H_F}$=
$-i(\sin\alpha \, \lambda_{3} v_{\rm SM} $+$ 2 \cos\alpha \, \lambda_{2} v_s)$=$i\,\cos\alpha\,(M_{{H_F}}^2+M_{{A_F}}^2)/v_s$. The effective coupling strengths with nucleons are given by $g_{N\overline{N} h}$
= $i(\cos\alpha \, \frac{M_N}{v_{\rm SM}} -2\sin\alpha \,Z_{ij}\, \frac{ v_{\rm SM}}{\sqrt{2} \,v_s} )$ and
$g_{N\overline{N} H_F} = i(\sin\alpha \, \frac{M_N}{v_{\rm SM}} +2\cos\alpha \,Z_{ij}\,  \frac{ v_{\rm SM}}{\sqrt{2} \,v_s} )$.
The $Z_{ij}$ can be found in the Eq.~(\ref{eq:fermiYukfinal}).
In the $t\to 0$ limit with $Z_{ij}=1$, the effective square DM–nucleon amplitude can be written as,
\begin{eqnarray}
    |\mathcal{M}_{\rm eff}|^2  &&= \frac{2}{ v_{\rm SM}^4 v_s^4 }\, [ v_s \sin2\alpha M_N (2 M_{A_F}^2+M_{H_F}^2+M_h^2)\nn\\
    &&~~~~~~~~~+2\sqrt{2} v_{\rm SM}^2 (\cos2\alpha M_{A_F}^2+\cos^2\alpha M_{H_F}^2-\sin^2\alpha M_h^2) ]^2,
\end{eqnarray}

One can always get $|\mathcal{M}_{\rm eff}|^2=\frac{16( M_{A_F}^2+ M_{H_F}^2)^2}{ v_s^4 }$ even for $\alpha \to 0$. It is to be noted that the mixing angle $\alpha$ strictly constrained by present Higgs signal strength data (see Figure~\ref{fig:DarkPlotkinmix}). Hence, in this model,
the tree-level direct-detection cross-section can not vanish
in the $t\to 0$ even after fine-tuning as in Refs~\cite{Alanne:2020jwx, Azevedo:2018exj}. 
By adjusting the coupling strengths parameters, one could get extreme fine-tuning, which simultaneously reduces the effective annihilation cross-section, leading to an overabundance of DM. In this scenario, the tree-level direct detection cross-section will always dominate the loop-level contributions (the details can be found in~\cite{Azevedo:2018exj}).
\bibliographystyle{utphys}
\bibliography{REFc}
\end{document}